\documentclass[journal]{IEEEtran}

\ifCLASSINFOpdf

\else

\fi

\usepackage{algpseudocode}
\usepackage{algorithm}
\usepackage[dvips]{graphicx}
\usepackage{latexsym}
\usepackage{amssymb}
\usepackage{amsmath}
\usepackage{multirow}
\usepackage{xcolor}
\usepackage{lipsum}
\usepackage{multicol}
\usepackage{subfig}
\usepackage{enumerate}
\usepackage{algorithm}
\usepackage{algorithmicx}
\usepackage{algpseudocode}
\usepackage{amsmath}
\usepackage{graphicx}
\usepackage[utf8]{inputenc}
\usepackage{float}
\usepackage{bm}

\begin{document}

\title{Practical Interference Exploitation Precoding without Symbol-by-Symbol Optimization: \\A Block-Level Approach}
\author{Ang Li,~\IEEEmembership{Senior Member,~IEEE}, Chao Shen,~\IEEEmembership{Member,~IEEE}, Xuewen Liao,~\IEEEmembership{Member,~IEEE}, \\ Christos Masouros,~\IEEEmembership{Senior Member,~IEEE}, and A. Lee Swindlehurst,~\IEEEmembership{Fellow,~IEEE}

\thanks{Manuscript received XX; revised XX. {\textit{(Corresponding author: Ang Li)}}}
\thanks{A. Li is with the School of Information and Communications Engineering, Faculty of Electronic and Information Engineering, Xi'an Jiaotong University, Xi'an, Shaanxi 710049, China, and is also with The State Key Laboratory of Integrated Services Networks, Xidian University, Xi’an, Shaanxi, China (e-mail: ang.li.2020@xjtu.edu.cn).}
\thanks{C. Shen is with the Shenzhen Research Institute of Big Data, Shenzhen 518172, China (e-mail: chaoshen@sribd.cn).}
\thanks{X. Liao is with the School of Information and Communications Engineering, Faculty of Electronic and Information Engineering, Xi'an Jiaotong University, Xi'an, Shaanxi 710049, China (e-mail: yeplos@mail.xjtu.edu.cn).}
\thanks{C. Masouros is with the Department of Electronic and Electrical Engineering, University College London, Torrington Place, London, WC1E 7JE, UK (e-mail: c.masouros@ucl.ac.uk).}
\thanks{A. L. Swindlehurst is with the Center for Pervasive Communications and Computing, Henry Samueli School of Engineering, University of California, Irvine, CA 92697, USA (e-mail: swindle@uci.edu).}
\thanks{This work was supported by xxx.}
}

\maketitle

\begin{abstract}
In this paper, we propose a constructive interference (CI)-based block-level precoding (CI-BLP) approach for the downlink of a multi-user multiple-input single-output (MU-MISO) communication system. Contrary to existing CI precoding approaches which have to be designed on a symbol-by-symbol level, here a constant precoding matrix is applied to a block of symbol slots within a channel coherence interval, thus significantly reducing the computational costs over traditional CI-based symbol-level precoding (CI-SLP) as the CI-BLP optimization problem only needs to be solved once per block. For both PSK and QAM modulation, we formulate an optimization problem to maximize the minimum CI effect over the block subject to a block- rather than symbol-level power budget. We mathematically derive the optimal precoding matrix for CI-BLP as a function of the Lagrange multipliers in closed form. By formulating the dual problem, the original CI-BLP optimization problem is further shown to be equivalent to a quadratic programming (QP) optimization. Numerical results validate our derivations, and show that the proposed CI-BLP scheme achieves improved performance over the traditional CI-SLP method, thanks to the relaxed power constraint over the considered block of symbol slots.
\end{abstract}


\IEEEpeerreviewmaketitle

\section{Introduction}
\IEEEPARstart{M}{ULTIPLE}-input multiple-output (MIMO) technology has been widely adopted in current cellular communication systems and will be an indispensable part of future wireless communication systems because of its significant gains over single-antenna systems \cite{r1}. In the downlink transmission of a multi-user MIMO communication system, precoding is essential for realizing spatial multiplexing. When channel state information (CSI) is available at the base station (BS), dirty-paper coding (DPC) is able to achieve the channel capacity by pre-subtracting the interference prior to transmission \cite{r2}. Despite its optimal performance, it is difficult to employ DPC in practical wireless systems due to its prohibitive computational costs and unrealistic assumption of an infinite alphabet. To alleviate the requirements of DPC, precoding approaches such as Tomlinson-Harashima precoding (THP) \cite{r3} and vector perturbation (VP) precoding \cite{r4} have been proposed. To further reduce the signal processing complexity, a number of closed-form linear precoding schemes have been studied, among which the most representative examples include zero-forcing (ZF) \cite{r5} and regularized ZF (RZF) precoding \cite{r6}. On the other hand, optimization-based precoding methods have received increasing research attention recently because of their flexibility in optimizing certain performance metrics \cite{r7}\nocite{r8}\nocite{r9}\nocite{r10}\nocite{r11}-\cite{WMMSE}. One popular form is downlink multicast precoding that targets broadcasting common information to all users \cite{r7}. Another popular example is the downlink signal-to-interference-plus-noise ratio (SINR) balancing approach, which aims to achieve a desired SINR for each user subject to either a total transmit power \cite{r8} or a per-antenna power budget \cite{r9}. An alternative optimization-based design aims to minimize the required transmit power at the BS for a given received SINR target for each user \cite{r10}. It is further shown in the literature that the power minimization and the SINR balancing problems are duals of one another \cite{r8}, \cite{r11}, and uplink-downlink duality can be exploited to obtain efficient iterative algorithms. In addition, the weighted minimum mean-squared error (W-MMSE) precoder was proposed in \cite{WMMSE} for weighted sum-rate maximization.

More recently, the concept of constructive interference (CI) has been introduced for downlink transmission in multi-user MIMO systems, and CI-based precoding has received increasing research attention. CI precoding is able to achieve improved performance over the above linear precoding methods by exploiting both the CSI and the data symbol information \cite{r12}, \cite{r13}. Unlike traditional practice where interference is treated as harmful to the system performance, the superiority of CI-based approaches lies in its recognition that, at the symbol level, the instantaneous interference can be categorized as either constructive or destructive interference (DI). This idea was first discussed in \cite{r14}, and a similar concept referred to as `convex vector precoding' was introduced in \cite{muller}. Based on this concept and by further exploiting the data symbol information in addition to CSI, a modified ZF precoder was designed in \cite{r15}, where the beneficial CI is preserved while DI is eliminated by the ZF process. A more advanced correlation rotation method was further proposed in \cite{r16}, in which a rotation matrix is applied to the precoder to modify all interference to be constructive.

As a step further, CI-based precoding has been implemented under various optimization criteria in order to achieve further performance improvement \cite{r17}\nocite{r18}\nocite{r19}\nocite{r20}-\cite{r21}. More specifically, \cite{r17} first proposed an optimization-based CI approach in the context of VP precoding by substituting the sophisticated sphere-search process with a linear scaling operation, leading to a quadratic programming (QP) formulation that reduces the computational complexity of traditional VP precoding. The work in \cite{r18} combines CI with maximum ratio transmission (MRT) precoding to improve the performance of the correlation-rotation CI precoding proposed in \cite{r16}. In addition, CI-based power minimization and SINR balancing problems are also studied. In \cite{r18},  the interfering signals in both cases are optimized to be strictly aligned with the intended data symbols to achieve CI, an approach that was later shown to be sub-optimal and referred to as the `strict phase-rotation' CI metric.

More advanced CI metrics are introduced in \cite{r19}, \cite{r20}, where the concept of the `constructive region' is introduced, within which all the interference is constructive to the intended data symbols. This observation alleviates the requirement that the interfering signals have to be strictly rotated to the direction of the intended data symbols, leading to further performance improvements. The CI metric introduced in \cite{r19} was later named the `non-strict phase-rotation' CI metric and is widely adopted in the relevant literature. Meanwhile, a relaxed CI metric based on a `relaxed detection region' was introduced in \cite{r20}, which expands the constructive region based on a phase margin that is related to the signal-to-noise ratio (SNR) target. The above CI-based precoding approaches \cite{r15}-\cite{r20} are all designed for PSK modulation, while \cite{r21} was the first to extend the exploitation of CI to QAM modulation, where the CI effect can be exploited by the outer constellation points of a QAM constellation by employing the `symbol-scaling' CI metric. Owing to the above benefits, the concept of CI has been applied to a number of wireless communication scenarios such as mutual coupling exploitation \cite{r22}, constant-envelope precoding \cite{r23}, \cite{lee-1}, 1-bit precoding \cite{r24}\nocite{r25}-\cite{r26}, radar-communication coexistence \cite{r27}, \cite{lee-2}, physical-layer security \cite{r28}, \cite{lee-3}, etc. For a more comprehensive literature review on CI and SLP, we refer the interested readers to \cite{r29} and \cite{r30}.

It has to be mentioned that the benefits of CI exploited by the above approaches come at the cost of symbol-by-symbol signal processing operations, i.e., symbol-level precoding (SLP), in which the precoder must be optimized on a symbol-by-symbol basis. This poses a significant computational burden on the multi-user MIMO communication system, because the BS will need to solve a different CI-SLP optimization problem for each symbol slot. To alleviate the computational costs, several studies attempt to reduce the complexity of the CI-SLP optimization problem, including derivations of the optimal precoding structure of CI-SLP with efficient iterative algorithms \cite{r31}, \cite{r32}, sub-optimal solutions \cite{r33}, \cite{r34}, and deep learning-based methods \cite{DL-1}\nocite{DL-2}-\cite{DL-3}. Specifically, \cite{r31} and \cite{r32} derive the optimal precoding structure of CI-SLP for PSK and QAM modulation, respectively, and show that the CI-SLP optimization problem can equivalently be transformed into a QP optimization problem and solved using an iterative algorithm with a closed-form solution at each step. The work in \cite{r33} derives an exact closed-form but sub-optimal solution for the power minimization CI-SLP problem, while \cite{r34} shows that CI-SLP precoding can be regarded as a symbol-level ZF precoding with a perturbation vector applied to the data symbols. Despite the above attempts to reduce the computational costs of solving the CI-SLP optimization problem for each symbol slot, all the above approaches still require solving an optimization problem at the symbol level, i.e., the total number of CI-SLP optimization problems that must be solved in a channel coherence interval is not reduced.

Therefore in this paper, for the first time in the literature we propose CI-based block-level precoding (CI-BLP) for both PSK and QAM modulation for the downlink of a MU-MISO system, which further motivates the use of CI-based precoding techniques in practical wireless communication systems. We summarize the main contributions of this paper below:

\begin{figure*}[!t]
\begin{centering}
\subfloat[Conventional BLP]
{\begin{centering}
\includegraphics[width=0.3\textwidth]{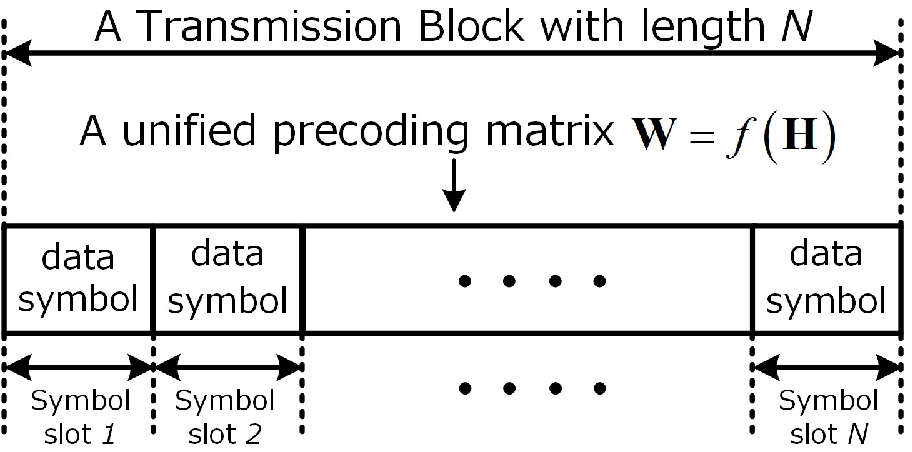}
\par
\end{centering}
}
\hspace{0.1cm}
\subfloat[Conventional CI-SLP]
{\begin{centering}
\includegraphics[width=0.3\textwidth]{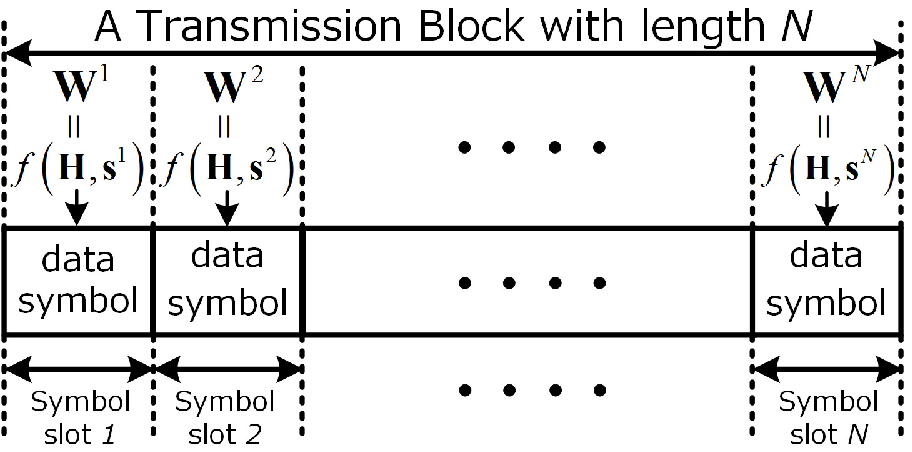}
\par
\end{centering}
}
\hspace{0.1cm}
\subfloat[Proposed CI-BLP]
{\begin{centering}
\includegraphics[width=0.3\textwidth]{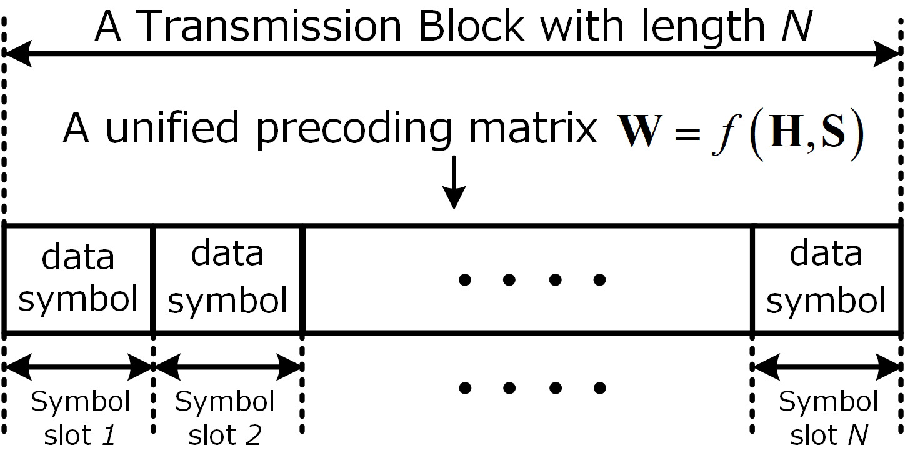}
\par
\end{centering}
}
\par
\end{centering}
\caption{A comparison between different precoding methodologies}
\end{figure*}

\begin{enumerate}

\item We propose CI-BLP that exploits CI on a block level for both PSK and QAM, where a constant precoding matrix is applied to a block of symbol slots within a channel coherence interval. The corresponding optimization problem is formulated to maximize the minimum CI effect over all symbol slots employing the `symbol-scaling' CI metric, subject to a block-level power budget.

\item For PSK modulation, based on the Lagrangian and Karush-Kuhn-Tucker (KKT) method, we derive the optimal precoding matrix for CI-BLP in closed form as a function of the Lagrange multipliers. By further studying the corresponding dual problem, the original CI-BLP optimization problem for PSK modulation is finally shown to be equivalent to a QP optimization over a simplex.

\item We further extend our mathematical analysis to QAM modulation, where we derive the optimal precoding matrix for CI-BLP and the dual problem formulation in a similar way. The original CI-BLP optimization problem for QAM modulation is also shown to be equivalent to a QP optimization, but not over a simplex any more. Another advantage of the proposed CI-BLP with QAM modulation is that it returns a constant power normalization factor over the considered block of symbol slots, thus reducing the signaling overhead for traditional CI-SLP with QAM.

\item Our above analyses for the PSK and QAM cases reveal that the proposed CI-BLP scheme shares a similar problem structure to the conventional CI-SLP method. While the problem size for the proposed CI-BLP approach is scaled up owing to the joint design over a block of symbol slots, the computational cost is reduced compared to traditional CI-SLP because the optimization problem only needs to be solved once per block.

\end{enumerate}

Numerical results validate our derivations, and show that 1) the proposed CI-BLP approach can achieve an improved performance over the conventional CI-SLP scheme when the length of the block is short and the transmit power is sufficiently high, thanks to the relaxed block-level power budget; 2) the proposed CI-BLP method only exhibits a slight performance loss compared to conventional CI-SLP as the length of the block increases; 3) the proposed CI-BLP scheme offers reduced computational costs compared to traditional CI-SLP methods, as validated by the execution time result.

The remainder of this paper is organized as follows. Section II introduces the system model and briefly reviews CI. Section III studies the proposed CI-BLP scheme for PSK modulation, and Section IV studies the proposed CI-BLP for QAM modulation. Numerical results are shown in Section V, and Section VI concludes the paper.

$\bf Notation:$ $a$, $\bf a$, and $\bf A$ denote scalar, column vector and matrix, respectively. ${( \cdot )^*}$, $( \cdot )^\text{T}$, and $( \cdot )^\text{H}$ denote conjugate, transposition, and conjugate transposition, respectively. ${\bf{A}}\left( {k,i} \right)$ denotes the entry in the $k$-row and $i$-th column of $\bf A$. ${{\mathbb C}^{n \times n}}$ (${{\mathbb R}^{n \times n}}$) represents an $n \times n$ matrix in the complex (real) set, and ${\bf I}_{K}$ denotes a $K \times K$ identity matrix. $\Re ( \cdot )$ and $\Im ( \cdot )$ extract the real and imaginary part of the argument, respectively. $\left\|  \cdot  \right\|_2$ denotes the $\ell_2$-norm, and $\jmath $ represents the imaginary unit. $\text{card}\left\{ \cdot\right\}$ is the cardinality of a set.

\section{System Model and Constructive Interference}
\subsection{System Model}
We consider a multi-user MISO (MU-MISO) system during downlink transmission, where a BS equipped with $N_\text{T}$ transmit antennas is communicating with a total number of $K$ single-antenna users in the same time-frequency resource, and where $K \le N_\text{T}$. We introduce ${\bf S}=\left[ {{\bf s}^1, {\bf s}^2, \cdots, {\bf s}^N } \right] \in {\mathbb C}^{K \times N}$ as the data symbol matrix for the considered block of symbol slots, where $N$ represents the length of the considered block which may be smaller than the channel coherence interval. The vector ${\bf s}^n = \left [ {s_1^n, s_2^n, \cdots, s_K^n} \right]^\text{T} \in {\mathbb C}^{K}$ contains the users' symbols for the $n$-th slot, drawn from normalized PSK or QAM constellations. Accordingly, the received signal for the $k$-th user in the $n$-th symbol slot is given by
\begin{equation}
y_k^n= {\bf h}_k^\text{T}{\bf W}{\bf s}^n + z_k^n,
\label{eq_1}
\end{equation}
where ${\bf h}_k \in {\mathbb C}^{N_\text{T}}$ represents the flat-fading channel vector between the BS and user $k$, which is constant within a channel coherence interval, and $z_k^n$ is additive Gaussian noise with zero mean and variance $\sigma^2$. The matrix ${\bf W} \in {\mathbb C}^{N_\text{T} \times K}$ is the block-level precoding matrix that is applied to all symbol slots in the block.

Fig. 1 depicts the difference between our proposed CI-BLP scheme and other standard precoding approaches, where $\bf H$ represents the CSI. Compared to existing CI-SLP approaches that optimize the precoding matrix (or the precoded signals) on a symbol level, our proposed CI-BLP scheme applies a constant precoding matrix $\bf W$ to all symbol slots, thus reducing the computational costs for the BS because the optimization only needs to be performed once per block of symbol slots. Moreover, our proposed CI-BLP approach is a linear precoding method, while CI-SLP has both linear and non-linear implementations. When compared to conventional BLP that also applies the same precoding matrix to all symbol slots based only on the CSI, our proposed CI-BLP scheme further exploits the data symbol information available at the BS for additional performance improvements, i.e., conventional BLP is linear data-independent precoding, while CI-BLP is linear data-dependent precoding. Since we focus on deriving the precoding structure for the proposed CI-BLP method, perfect CSI is assumed throughout the paper.

\subsection{Constructive Interference}
CI is defined as interference that is able to push the received signals away from the corresponding decision boundaries of the modulated symbol constellation, and the constructive region is defined as the area within which the received signals enjoy a larger distance to the decision boundaries compared to the corresponding nominal constellation point \cite{r29}. In this paper, the `symbol-scaling' CI metric is employed for both PSK and QAM modulation. As an illustrative example, Fig. 2 presents one quarter of a 8PSK constellation, where the green shaded area represents the constructive region corresponding to the 8PSK constellation point in the first quadrant. Without loss of generality, we assume $\vec {OS}=s_k^n$ is the data symbol of interest for user $k$ in the $n$-th symbol slot, and we introduce $\vec{OB}={\bf h}_k^\text{T}{\bf W}{\bf s}^n$ as the corresponding received signal for user $k$ excluding noise. From this perspective, the effect of interference is equivalent to a scaling and rotation operation on the transmitted data symbol $\vec{OS}$. Based on the geometry, if node `B' is located in the constructive region, as depicted in Fig. 2, the distance between node `B' and the two decision boundaries $\vec{OP}$ and $\vec{OQ}$ is larger than that between the nominal constellation point `S' and the two decision boundaries, leading to an effective increase in the received SINR and a reduction in the probability of a detection error. Accordingly, CI is achieved in this case because the effect of interference contributes to the useful signal power. For a more detailed discussion on different CI metrics, we refer the interested readers to \cite{r29}.

\begin{figure}[!t]
\centering
\includegraphics[scale=0.35]{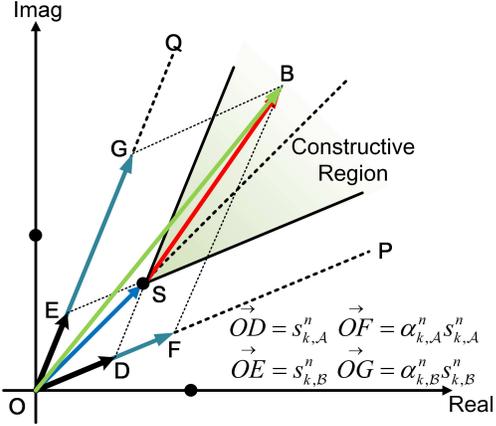}
\caption{An illustration for CI, 8PSK, `symbol-scaling' metric}
\end{figure}

\newcounter{mytempeqncnt1}
\begin{figure*}[!t]
\normalsize
\setcounter{equation}{13}

\begin{equation}
\begin{aligned}
&{\cal L}_1\left ( {\bf \hat W} , {t}, {\delta_{k}^{n}}, {\mu}  \right ) \\
& = -{t} +\sum_{n=1}^{N} \sum_{k=1}^{2K} {\delta_k^n}\left [ {{t}- \left( {{\bf a}_k^n} \right)^\text{T}{\bf \hat W} {\bf s}_\text{E}^n - \left( {{\bf b}_k^n} \right)^\text{T}{\bf \hat W} {\bf c}_\text{E}^n} \right ] + {\mu} \left[{\sum_{n=1}^{N} \left ( {\bf s}_\text{E}^n \right )^\text{T} \left ( {{\bf P} {\bf \hat W} + {\bf Q} {\bf \hat W} {\bf T}} \right )^\text{T}\left ( {{\bf P} {\bf \hat W} + {\bf Q} {\bf \hat W} {\bf T}} \right ) {\bf s}_\text{E}^n -N p_0 }\right] \\
&=\left ( {\sum_{n=1}^{N} {\bf 1}^\text{T}{\bm{\delta}}^n -1 } \right ) {t} -\sum_{n=1}^{N}\left( {{\bm{ \delta}}^n} \right)^\text{T}{\bf A}^n{\bf \hat W}{\bf s}_\text{E}^n -\sum_{n=1}^{N}\left( {{\bm{\delta}}^n} \right)^\text{T}{\bf B}^n{\bf \hat W}{\bf c}_\text{E}^n \\
& {\kern 10pt} + {\mu} \left[{\sum_{n=1}^{N} \left ( {\bf s}_\text{E}^n \right )^\text{T} \left ( { {\bf \hat W}^\text{T} {\bf P}^\text{T}  + {\bf T}^\text{T} {\bf \hat W}^\text{T}  {\bf Q}^\text{T}  } \right )\left ( {{\bf P} {\bf \hat W} + {\bf Q} {\bf \hat W} {\bf T}} \right ) {\bf s}_\text{E}^n }\right] - {\mu}N p_0\\
&=\left ( {\sum_{n=1}^{N} {\bf 1}^\text{T}{\bm{\delta}}^n -1 } \right ) {t} -\sum_{n=1}^{N}\left( {{\bm{ \delta}}^n} \right)^\text{T}{\bf A}^n{\bf \hat W}{\bf s}_\text{E}^n -\sum_{n=1}^{N}\left( {{\bm{\delta}}^n} \right)^\text{T}{\bf B}^n{\bf \hat W}{\bf c}_\text{E}^n + {\mu}\sum_{n=1}^{N} \left ( {\bf s}_\text{E}^n  \right )^\text{T} {\bf \hat W}^\text{T} {\bf P}^\text{T}{\bf P}{\bf \hat W} {\bf s}_\text{E}^n  \\
& {\kern 10pt} + \underbrace{{\mu}\sum_{n=1}^{N} \left ( {\bf s}_\text{E}^n  \right )^\text{T} {\bf \hat W}^\text{T} {\bf P}^\text{T}{\bf Q}{\bf \hat W}{\bf T}{\bf s}_\text{E}^n}_{=0}  + \underbrace{{\mu}\sum_{n=1}^{N} \left ( {\bf s}_\text{E}^n  \right )^\text{T} {\bf T}^\text{T} {\bf \hat W}^\text{T}  {\bf Q}^\text{T} {\bf P} {\bf \hat W} {\bf s}_\text{E}^n}_{=0}  + {\mu}\sum_{n=1}^{N} \underbrace{\left ( {\bf s}_\text{E}^n  \right )^\text{T} {\bf T}^\text{T}}_{\left( {\bf c}_\text{E}^n  \right)^\text{T} }  {\bf \hat W}^\text{T}  {\bf Q}^\text{T}{\bf Q}{\bf \hat W}\underbrace{{\bf T}{\bf s}_\text{E}^n}_{{\bf c}_\text{E}^n}  -{u_0} N p_0 \\
&=\left ( {\sum_{n=1}^{N} {\bf 1}^\text{T}{\bm{\delta}}^n -1 } \right ) {t} -\sum_{n=1}^{N}\left( {{\bm{ \delta}}^n} \right)^\text{T}{\bf A}^n{\bf \hat W}{\bf s}_\text{E}^n -\sum_{n=1}^{N}\left( {{\bm{\delta}}^n} \right)^\text{T}{\bf B}^n{\bf \hat W}{\bf c}_\text{E}^n + {\mu}\sum_{n=1}^{N} \left ( {\bf s}_\text{E}^n  \right )^\text{T} {\bf \hat W}^\text{T} {\bf \hat W} {\bf s}_\text{E}^n + {\mu}\sum_{n=1}^{N} \left ( {\bf c}_\text{E}^n  \right )^\text{T} {\bf \hat W}^\text{T} {\bf \hat W} {\bf c}_\text{E}^n -{u_0}Np_0
\end{aligned}
\label{eq_14}
\end{equation}

\hrulefill
\vspace*{4pt}
\end{figure*}

\newcounter{mytempeqncnt2}
\begin{figure*}[!t]
\normalsize
\setcounter{equation}{15}

\begin{IEEEeqnarray}{rCl}
\IEEEyesnumber
\frac{{\partial {\cal L}_1}}{{\partial t}} =\sum_{n=1}^{N} {\bf 1}^\text{T}{\bm{\delta}}^n -1 =0  {\kern 30pt} \IEEEyessubnumber* \label{eq_16a} \\
\frac{{\partial {\cal L}_1}}{{\partial {\bf \hat W}}} = -\sum_{n=1}^{N} \left [ {\left ( {{\bm{\delta}}^n} \right ) }^\text{T} {\bf A}^n \right ] ^\text{T} \left ( {\bf s}_\text{E}^n  \right )^\text{T} -\sum_{n=1}^{N} \left [ {\left ( {{\bm{\delta}}^n} \right ) }^\text{T} {\bf B}^n \right ] ^\text{T} \left ( {\bf c}_\text{E}^n  \right )^\text{T} + 2{u_0} {\bf \hat W}\left [ {\sum_{n=1}^{N} {\bf s}_\text{E}^n  \left ( {\bf s}_\text{E}^n \right )^\text{T} + \sum_{n=1}^{N} {\bf c}_\text{E}^n  \left ( {\bf c}_\text{E}^n \right )^\text{T} }  \right ]  = {\bf 0}  {\kern 30pt} \label{eq_16b} \\
{\delta_k^n}\left [ {{t}- \left( {{\bf a}_k^n} \right)^\text{T}{\bf \hat W} {\bf s}_\text{E}^n - \left( {{\bf b}_k^n} \right)^\text{T}{\bf \hat W} {\bf c}_\text{E}^n} \right ]=0, {\kern 3pt} {\delta_k^n} \ge 0, {\kern 3pt} \forall k \in {\cal K}, {\kern 3pt} \forall n \in {\cal N} {\kern 30pt} \label{eq_16c} \\
{\mu}\left [ \sum_{n=1}^{N} \left ( {\bf s}_\text{E}^n  \right )^\text{T} {\bf \hat W}^\text{T} {\bf \hat W} {\bf s}_\text{E}^n + \sum_{n=1}^{N} \left ( {\bf c}_\text{E}^n  \right )^\text{T} {\bf \hat W}^\text{T} {\bf \hat W} {\bf c}_\text{E}^n -Np_0 \right ] =0, {\kern 3pt} {\mu} \ge 0 {\kern 30pt} \label{eq_16d}
\end{IEEEeqnarray}

\hrulefill
\vspace*{4pt}
\end{figure*}

\section{Proposed CI-BLP for PSK Modulation}
\subsection{Problem Formulation}
In this section, we study the proposed CI-BLP optimization for PSK modulation. Since it has already been shown in \cite{r32} that the `non-strict phase-rotation' CI metric commonly adopted for PSK modulation is equivalent to the `symbol-scaling' CI metric for PSK, we employ the `symbol-scaling' CI metric in this section to be consistent with the mathematics of Section IV.

The `symbol-scaling' CI metric decomposes both the modulated data symbols and the noiseless received signals along their corresponding decision boundaries, where two real-valued scaling coefficients are introduced to the noiseless received signals to jointly represent the scaling and rotation effect of the interference on the intended symbol of interest. Based on this criterion, to apply the `symbol-scaling' CI metric to PSK modulation, each desired data symbol $\vec{OS}=s_k^n$ and the corresponding received signal excluding noise $\vec{OB}={\bf h}_k^\text{T}{\bf W}{\bf s}^n$ are decomposed along the two decision boundaries corresponding to $s_k^n$, as depicted in Fig. 2. Mathematically, this decomposition can be expressed as
\begin{equation}
\begin{aligned}
\setcounter{equation}{2}
 \vec{OS}&=\vec{OD}+\vec{OE} {\kern 1pt} \Rightarrow {\kern 1pt} s_k^n = s_{k, \cal A}^n +  s_{k, \cal B}^n, \\
 \vec{OB}&=\vec{OF}+\vec{OG} {\kern 2pt} \Rightarrow {\kern 1pt} {\bf h}_k^\text{T}{\bf W}{\bf s}^n =\alpha_{k, \cal A}^n s_{k, \cal A}^n + \alpha_{k, \cal B}^n s_{k, \cal B}^n,
\end{aligned}
\label{eq_2}
\end{equation}
where $\alpha_{k, \cal A}^n \ge 0$ and $\alpha_{k, \cal B}^n \ge 0$ are real-valued scaling coefficients. We observe that a larger value of $\alpha_{k, \cal A}^n$ or $\alpha_{k, \cal B}^n$ represents a larger distance to the decision boundaries, and therefore a larger CI effect and a better error rate performance. We define ${\bm \alpha}_\text{E}^n \in {\mathbb R}^{2K}$ as
\begin{equation}
{\bm \alpha}_\text{E}^n=\left[ {\alpha_{1, \cal A}^n, \alpha_{2, \cal A}^n, \cdots, \alpha_{K, \cal A}^n, \alpha_{1, \cal B}^n, \alpha_{2, \cal B}^n, \cdots, \alpha_{K, \cal B}^n} \right] ^\text{T},
\label{eq_3}
\end{equation}
and by following the transformations in \cite{r25}, ${\bm \alpha}_\text{E}^n$ can be further expressed as
\begin{equation}
{\bm \alpha}_\text{E}^n={\bf M}^n {\bf W}_\text{E} {\bf s}_\text{E}^n,
\label{eq_4}
\end{equation}
where the construction of ${\bf M}^n \in {\mathbb R}^{2K \times 2N_\text{T}}$ is shown in Appendix A, which directly follows Section IV-A of \cite{r25}. ${\bf W}_\text{E} \in {\mathbb R}^{2N_\text{T} \times 2K}$ and ${\bf s}_\text{E}^n \in {\mathbb R}^{2K}$ in \eqref{eq_4} are defined as
\begin{equation}
{\bf W}_\text{E} = \left [ \begin{matrix}
 \Re \left ( \bf W \right )  & - \Im \left ( \bf W \right ) \\
 \Im \left ( \bf W \right ) & \Re \left ( \bf W \right )
\end{matrix} \right ], {\kern 3pt} {\bf s}_\text{E}^n= \left[ {\Re \left( {{\bf s}^n} \right)^\text{T}, \Im \left( {{\bf s}^n} \right)^\text{T} } \right]^\text{T}.
\label{eq_5}
\end{equation}

The proposed CI-BLP optimization aims to maximize the minimum entry in ${\bm \alpha}_\text{E}^n$ for all symbol slots within the considered transmission block, and the corresponding optimization problem can thus be constructed as:
\begin{equation}
\begin{aligned}
&\mathcal{P}_0^\text{PSK}: {\kern 3pt} \max_{{\bf W}_\text{E}} \min_{k,n} {\kern 3pt} \alpha_k^n \\
&{\kern 6pt} \text{s.t.} {\kern 13pt} {\bf C1:} {\kern 3pt} {\bm \alpha}_\text{E}^n={\bf M}^n {\bf W}_\text{E} {\bf s}_\text{E}^n, {\kern 3pt} \forall n \in {\cal N}, \\
&{\kern 31pt} {\bf C2:} {\kern 3pt} \sum_{n=1}^{N} \left \| {{\bf W}_\text{E} {\bf s}_\text{E}^n  }  \right \| _2^2 \le N{p_0},
\label{eq_6}
\end{aligned}
\end{equation}
where $\alpha_k^n$ represents the $k$-th entry in ${\bm \alpha}_\text{E}^n$, $p_0$ represents the transmit power budget per symbol slot, and ${\cal N}=\left\{ {1,2,\cdots,N} \right\}$. Compared to the traditional CI-SLP problem, $\mathcal{P}_0^\text{PSK}$ differs in that

\begin{enumerate}
\item a constant precoding matrix is applied to all symbols in the considered block, and it is optimized across all symbol slots, thus greatly reducing the computational costs over the traditional CI-SLP approach;
\item the power budget is enforced over the entire considered block instead of within each symbol slot, i.e., a relaxed power constraint is enforced compared to traditional CI-SLP.
\end{enumerate}
The relaxed power constraint is an advantage for CI-BLP, but the use of a constant precoder for all symbols is a disadvantage compared with CI-SLP. For small $N$ and a large enough transmit power, the benefit of the relaxed power constraint outweighs the loss due to using a fixed precoder, leading to a better performance for CI-BLP over CI-SLP.

$\mathcal{P}_0^\text{PSK}$ is a joint optimization over all symbol slots, and it is a convex problem that can be directly solved via optimization tools such as CVX. To facilitate subsequent derivations, we introduce $\bf \hat W$:
\begin{equation}
{\bf \hat W} = \begin{bmatrix}
 \Re \left ( \bf W \right ) & -\Im \left ( \bf W \right )
\end{bmatrix} \in {\mathbb R}^{N_\text{T} \times 2K},
\label{eq_7}
\end{equation}
based on which we can decompose ${\bf W}_\text{E}$ into
\begin{equation}
{\bf W}_\text{E}  = {\bf P} {\bf \hat W} + {\bf Q} {\bf \hat W} {\bf T},
\label{eq_8}
\end{equation}
where ${\bf P} \in {\mathbb R}^{2N_\text{T} \times N_\text{T}}$, ${\bf Q} \in {\mathbb R}^{2N_\text{T} \times N_\text{T}}$ and ${\bf T} \in {\mathbb R}^{2K \times 2K}$ are defined as
\begin{equation}
{\bf P}= \left [ \begin{matrix}
{\bf I}_{N_\text{T}}  \\
{\bf 0}
\end{matrix} \right ], {\kern 3pt} {\bf Q}=\left [ \begin{matrix}
{\bf 0}  \\
{\bf I}_{N_\text{T}}
\end{matrix} \right ], {\kern 3pt} {\bf T}= \left [ \begin{matrix}
 {\bf 0} & {\bf I}_K \\
 -{\bf I}_K & {\bf 0}
\end{matrix} \right ].
\label{eq_9}
\end{equation}
Based on \eqref{eq_8}, the expression for ${\bm \alpha}_\text{E}^n$ is further transformed into:
\begin{equation}
\begin{aligned}
{\bm \alpha}_\text{E}^n &= {\bf M}^n {\bf W}_\text{E} {\bf s}_\text{E}^n \\
&={\bf M}^n \left( {{\bf P} {\bf \hat W} + {\bf Q} {\bf \hat W} {\bf T}} \right) {\bf s}_\text{E}^n \\
&={\bf M}^n {\bf P} {\bf \hat W} {\bf s}_\text{E}^n + {\bf M}^n {\bf Q} {\bf \hat W} {\bf T} {\bf s}_\text{E}^n\\
&={\bf A}^n{\bf \hat W} {\bf s}_\text{E}^n + {\bf B}^n {\bf \hat W} {\bf c}_\text{E}^n,
\end{aligned}
\label{eq_10}
\end{equation}
where we introduce ${\bf A}^n \in {\mathbb R}^{2K \times N_\text{T}}$, ${\bf B}^n \in {\mathbb R}^{2K \times N_\text{T}}$ and ${\bf c}_\text{E}^n \in {\mathbb R}^{2K}$ as
\begin{equation}
{\bf A}^n= {\bf M}^n {\bf P}, {\kern 3pt} {\bf B}^n={\bf M}^n {\bf Q}, {\kern 3pt} {\bf c}_\text{E}^n={\bf T} {\bf s}_\text{E}^n.
\label{eq_11}
\end{equation}
With the expression for the $k$-th entry of ${\bm \alpha}_\text{E}^n$ given by
\begin{equation}
\alpha_k^n = \left( {{\bf a}_k^n} \right)^\text{T}{\bf \hat W} {\bf s}_\text{E}^n + \left( {{\bf b}_k^n} \right)^\text{T}{\bf \hat W} {\bf c}_\text{E}^n,
\label{eq_12}
\end{equation}
$\mathcal{P}_0^\text{PSK}$ can be expressed in the form of the standard convex optimization problem below:
\begin{equation}
\begin{aligned}
&\mathcal{P}_1^\text{PSK}: {\kern 3pt} \min_{{\bf \hat W}, {t}} - {t} \\
&\text{s.t.} {\kern 1pt} {\bf C1:} {\kern 2pt}  {t} - \left( {{\bf a}_k^n} \right)^\text{T}{\bf \hat W} {\bf s}_\text{E}^n - \left( {{\bf b}_k^n} \right)^\text{T}{\bf \hat W} {\bf c}_\text{E}^n \le 0, {\kern 1pt} \forall k \in {\cal K}, n \in {\cal N}, \\
&{\kern 12.5pt} {\bf C2:} {\kern 2pt}  \sum_{n=1}^{N} \left \| {\left({{\bf P} {\bf \hat W} + {\bf Q} {\bf \hat W} {\bf T}}\right) {\bf s}_\text{E}^n  }  \right \| _2^2 - N{p_0} \le 0,
\label{eq_13}
\end{aligned}
\end{equation}
where ${\cal K} = \left\{ {1,2,\cdots,2K} \right\}$.

\newcounter{mytempeqncnt3}
\begin{figure*}[!t]
\normalsize
\setcounter{equation}{21}

\begin{equation}
\begin{aligned}
{\cal U}_1 & = \max_{\left\{{{\bm{\delta}}^m}\right\}, {\mu}} -\sum_{m=1}^{N} \left ( {{\bm{\delta}}^m} \right )^\text{T}{\bf A}^m{\bf \hat W}{\bf s}_\text{E}^m - \sum_{m=1}^{N} \left ( {{\bm{\delta}}^m} \right )^\text{T}{\bf B}^m{\bf \hat W}{\bf c}_\text{E}^m \\
& = \min_{\left\{{{\bm{\delta}}^m}\right\}, {\mu}} \sum_{m=1}^{N} \left ( {{\bm{\delta}}^m} \right )^\text{T}{\bf A}^m  \frac{1}{2{\mu}} \sum_{n=1}^{N} \left [ {\left ( {\bf A}^n \right ) }^\text{T} {{\bm{\delta}}^n} \left ( {\bf s}_\text{E}^n  \right )^\text{T} + {\left ( {\bf B}^n \right ) }^\text{T} {{\bm{\delta}}^n} \left ( {\bf c}_\text{E}^n  \right )^\text{T} \right ] {\bf D}^{-1} {\bf s}_\text{E}^m \\
& {\kern 36pt} + \sum_{m=1}^{N} \left ( {{\bm{\delta}}^m} \right )^\text{T}{\bf B}^m \frac{1}{2{\mu}} \sum_{n=1}^{N} \left [ {\left ( {\bf A}^n \right ) }^\text{T} {{\bm{\delta}}^n} \left ( {\bf s}_\text{E}^n  \right )^\text{T} + {\left ( {\bf B}^n \right ) }^\text{T} {{\bm{\delta}}^n} \left ( {\bf c}_\text{E}^n  \right )^\text{T} \right ] {\bf D}^{-1} {\bf c}_\text{E}^m \\
& = \min_{\left\{{{\bm{\delta}}^m}\right\}, {\mu}} {\kern 1pt} \frac{1}{2 \mu} \sum_{m=1}^{N} \sum_{n=1}^{N} \left ( {{\bm{ \delta}}^m} \right ) ^\text{T}{\bf A}^m \left ( {\bf A}^n \right )^\text{T}{\bm{\delta}}^n \left ( {{\bf s}_\text{E}^n} \right )^\text{T}{\bf D}^{-1} {{\bf s}_\text{E}^m} + \frac{1}{2 \mu} \sum_{m=1}^{N} \sum_{n=1}^{N} \left ( {{\bm{\delta}}^m} \right ) ^\text{T}{\bf A}^m \left ( {\bf B}^n \right )^\text{T}{\bm{\delta}}^n \left ( {{\bf c}_\text{E}^n} \right )^\text{T}{\bf D}^{-1} {{\bf s}_\text{E}^m}   \\
& {\kern 36pt} + \frac{1}{2 \mu} \sum_{m=1}^{N} \sum_{n=1}^{N} \left ( {{\bm{\delta}}^m} \right ) ^\text{T}{\bf B}^m \left ( {\bf A}^n \right )^\text{T}{\bm{\delta}}^n \left ( {{\bf s}_\text{E}^n} \right )^\text{T}{\bf D}^{-1} {{\bf c}_\text{E}^m} + \frac{1}{2 \mu} \sum_{m=1}^{N} \sum_{n=1}^{N} \left ( {{\bm{\delta}}^m} \right ) ^\text{T}{\bf B}^m \left ( {\bf B}^n \right )^\text{T}{\bm{\delta}}^n \left ( {{\bf c}_\text{E}^n} \right )^\text{T}{\bf D}^{-1} {{\bf c}_\text{E}^m}
\end{aligned}
\label{eq_22}
\end{equation}

\hrulefill
\vspace*{4pt}
\end{figure*}

\newcounter{mytempeqncnt4}
\begin{figure*}[!t]
\normalsize
\setcounter{equation}{23}

\begin{equation}
\begin{aligned}
{\cal U}_1 &=  \min_{\left\{{{\bm{\delta}}^m}\right\}, {\mu}} \frac{1}{2 \mu} \sum_{m=1}^{N} \sum_{n=1}^{N} \left ( {{\bm{\delta}}^m} \right ) ^\text{T}\left [ p_{m,n} {\bf A}^m \left ( {\bf A}^n \right )^\text{T} \right ] {\bm{\delta}}^n + \frac{1}{2 \mu} \sum_{m=1}^{N} \sum_{n=1}^{N} \left ( {{\bm{\delta}}^m} \right ) ^\text{T}\left [ f_{m,n} {\bf A}^m \left ( {\bf B}^n \right )^\text{T} \right ] {\bm{\delta}}^n  \\
&{\kern 35pt} +\frac{1}{2 \mu} \sum_{m=1}^{N} \sum_{n=1}^{N} \left ( {{\bm{\delta}}^m} \right ) ^\text{T}\left [ g_{m,n} {\bf B}^m \left ( {\bf A}^n \right )^\text{T} \right ] {\bm{\delta}}^n +\frac{1}{2 \mu} \sum_{m=1}^{N} \sum_{n=1}^{N} \left ( {{\bm{\delta}}^m} \right ) ^\text{T}\left [ q_{m,n} {\bf B}^m \left ( {\bf B}^n \right )^\text{T} \right ] {\bm{\delta}}^n \\
&= \min_{\left\{{{\bm{\delta}}^m}\right\}, {\mu}} \frac{1}{2 \mu} \sum_{m=1}^{N} \sum_{n=1}^{N} \left ( {{\bm{\delta}}^m} \right ) ^\text{T}\left [ p_{m,n} {\bf A}^m \left ( {\bf A}^n \right )^\text{T} + f_{m,n} {\bf A}^m \left ( {\bf B}^n \right )^\text{T} + g_{m,n} {\bf B}^m \left ( {\bf A}^n \right )^\text{T} + q_{m,n} {\bf B}^m \left ( {\bf B}^n \right )^\text{T} \right ] {\bm{\delta}}^n
\end{aligned}
\label{eq_24}
\end{equation}

\hrulefill
\vspace*{4pt}
\end{figure*}

\subsection{Optimal Closed-Form Structure for $\bf \hat W$}
We derive the optimal precoding matrix $\bf \hat W$ based on the Lagrangian and KKT conditions. To begin with, the Lagrangian of $\mathcal{P}_1^\text{PSK}$ can be constructed as shown in \eqref{eq_14} on the top of this page, where ${\bm{\delta}}^n = \left[ \delta_1^n, \delta_2^n, \cdots, \delta_{2K}^n \right]^\text{T} \in {\mathbb R}^{2K}$ and $\mu$ are the non-negative dual variables associated with the inequality constraints ${\bf C1}$ and ${\bf C2}$ respectively, and the last step is obtained by the fact that
\begin{equation}
\setcounter{equation}{15}
{\bf P}^\text{T}{\bf P}={\bf Q}^\text{T}{\bf Q} = {\bf I}_{N_\text{T}}, {\kern 3pt} {\bf P}^\text{T}{\bf Q} = {\bf Q}^\text{T}{\bf P} = {\bf 0}.
\label{eq_15}
\end{equation}
Accordingly, the KKT conditions for optimality of $\mathcal{P}_1^\text{PSK}$ can be formulated and are shown in (16) above. Based on the KKT conditions, we first obtain that $\mu >0$, otherwise ${\bm \delta}^n={\bf 0}$, $\forall n$, which contradicts \eqref{eq_16a}, meaning that the block-level power constraint is active when optimality is achieved, i.e.,
\begin{equation}
\setcounter{equation}{17}
 \sum_{n=1}^{N} \left ( {\bf s}_\text{E}^n  \right )^\text{T} {\bf \hat W}^\text{T} {\bf \hat W} {\bf s}_\text{E}^n + \sum_{n=1}^{N} \left ( {\bf c}_\text{E}^n  \right )^\text{T} {\bf \hat W}^\text{T} {\bf \hat W} {\bf c}_\text{E}^n = Np_0.
\label{eq_17}
\end{equation}
This is also intuitive because a linear scaling to $\bf \hat W$ directly leads to an increase in the objective value $t$. To proceed, we transform \eqref{eq_16b} into
\begin{equation}
2{\mu} {\bf \hat W} {\bf D} = \sum_{n=1}^{N} \left [ {\left ( {\bf A}^n \right ) }^\text{T} {{\bm{\delta}}^n} \left ( {\bf s}_\text{E}^n  \right )^\text{T} + {\left ( {\bf B}^n \right ) }^\text{T} {{\bm{\delta}}^n} \left ( {\bf c}_\text{E}^n  \right )^\text{T} \right ],
\label{eq_18}
\end{equation}
where ${\bf D} \in {\mathbb R}^{2K \times 2K}$ is given by
\begin{equation}
{\bf D} = \left [ {\sum_{n=1}^{N} {\bf s}_\text{E}^n  \left ( {\bf s}_\text{E}^n \right )^\text{T} + \sum_{n=1}^{N} {\bf c}_\text{E}^n  \left ( {\bf c}_\text{E}^n \right )^\text{T} }  \right ].
\label{eq_19}
\end{equation}
Based on the fact that the block length $N$ is in general larger than the number of users $K$, $\bf D$ is thus full-rank and invertible \cite{3GPP}. Accordingly, we can obtain an expression for the optimal precoding matrix ${\bf \hat W}$ as a function of the Lagrange multipliers ${\bm \delta}^n$ in closed form as
\begin{equation}
{\bf \hat W} = \frac{1}{2{\mu}} \sum_{n=1}^{N} \left [ {\left ( {\bf A}^n \right ) }^\text{T} {{\bm{\delta}}^n} \left ( {\bf s}_\text{E}^n  \right )^\text{T} + {\left ( {\bf B}^n \right ) }^\text{T} {{\bm{\delta}}^n} \left ( {\bf c}_\text{E}^n  \right )^\text{T} \right ] {\bf D}^{-1}.
\label{eq_20}
\end{equation}
We observe from \eqref{eq_20} that the expression for the optimal precoding matrix $\bf \hat W$ includes all the data symbols transmitted within the considered block. Moreover, the optimization on $\bf \hat W$ is now transformed into an optimization on the Lagrange multipliers ${\bm \delta}^n$. In what follows, we consider the dual problem of $\mathcal{P}_1^\text{PSK}$ to further simplify the proposed CI-BLP optimization problem.

\newcounter{mytempeqncnt5}
\begin{figure*}[!t]
\normalsize
\setcounter{equation}{28}

\begin{equation}
\begin{aligned}
&\sum_{l=1}^{N} \left ( {\bf s}_\text{E}^l  \right )^\text{T} {\bf \hat W}^\text{T} {\bf \hat W} {\bf s}_\text{E}^l \\
=& \frac{1}{4{\mu^2}} \sum_{l=1}^{N} \left ( {\bf s}_\text{E}^l  \right )^\text{T} \left \{ \sum_{m=1}^{N} \left [ \left ( {\bf A}^m \right )^\text{T}{\bm{\delta}}^m \left ( {\bf s}_\text{E}^m \right )^\text{T} + \left ( {\bf B}^m \right )^\text{T}{\bm{\delta}}^m \left ( {\bf c}_\text{E}^m \right )^\text{T} \right ] {\bf D}^{-1}  \right \}^\text{T} \left \{ \sum_{n=1}^{N} \left [ \left ( {\bf A}^n \right )^\text{T}{\bm{ \delta}}^n \left ( {\bf s}_\text{E}^n \right )^\text{T} + \left ( {\bf B}^n \right )^\text{T}{\bm{\delta}}^n \left ( {\bf c}_\text{E}^n \right )^\text{T} \right ] {\bf D}^{-1}  \right \} {\bf s}_\text{E}^l \\
=& \frac{1}{4{\mu^2}} \sum_{l=1}^{N} \left ( {\bf s}_\text{E}^l  \right )^\text{T} {\bf D}^{-1} \left \{ \sum_{m=1}^{N} \left [ {\bf s}_\text{E}^m \left ( {{\bm{\delta}}^m} \right )^\text{T} {\bf A}^m + {\bf c}_\text{E}^m \left ( {{\bm{\delta}}^m} \right )^\text{T} {\bf B}^m \right ]  \right \} \left \{ \sum_{n=1}^{N} \left [ \left ( {\bf A}^n \right )^\text{T}{\bm{\delta}}^n \left ( {\bf s}_\text{E}^n \right )^\text{T} + \left ( {\bf B}^n \right )^\text{T}{\bm{\delta}}^n \left ( {\bf c}_\text{E}^n \right )^\text{T} \right ]  \right \} {\bf D}^{-1} {\bf s}_\text{E}^l \\
=& \frac{1}{4{\mu^2}}\sum_{l=1}^{N}\sum_{m=1}^{N}\sum_{n=1}^{N} \left [ \underbrace{\left ( {\bf s}_\text{E}^l  \right )^\text{T} {\bf D}^{-1} {\bf s}_\text{E}^m}_{p_{m,l}}  \left ( {\bm{\delta}}^m \right )^\text{T} {\bf A}^m \left ( {\bf A}^n \right )^\text{T}{\bm{\delta}}^n \underbrace{\left ( {\bf s}_\text{E}^n  \right )^\text{T} {\bf D}^{-1} {\bf s}_\text{E}^l}_{p_{l,n}}  \right ]   \\
& + \frac{1}{4{\mu^2}}\sum_{l=1}^{N}\sum_{m=1}^{N}\sum_{n=1}^{N} \left [ \underbrace{\left ( {\bf s}_\text{E}^l  \right )^\text{T} {\bf D}^{-1} {\bf s}_\text{E}^m}_{p_{m,l}}  \left ( {\bm{\delta}}^m \right )^\text{T} {\bf A}^m \left ( {\bf B}^n \right )^\text{T}{\bm{\delta}}^n \underbrace{\left ( {\bf c}_\text{E}^n  \right )^\text{T} {\bf D}^{-1} {\bf s}_\text{E}^l}_{f_{l,n}}  \right ]   \\
& + \frac{1}{4{\mu^2}}\sum_{l=1}^{N}\sum_{m=1}^{N}\sum_{n=1}^{N} \left [ \underbrace{\left ( {\bf s}_\text{E}^l  \right )^\text{T} {\bf D}^{-1} {\bf c}_\text{E}^m}_{g_{m,l}}  \left ( {\bm{\delta}}^m \right )^\text{T} {\bf B}^m \left ( {\bf A}^n \right )^\text{T}{\bm{\delta}}^n \underbrace{\left ( {\bf s}_\text{E}^n  \right )^\text{T} {\bf D}^{-1} {\bf s}_\text{E}^l}_{p_{l,n}}  \right ]   \\
& + \frac{1}{4{\mu^2}}\sum_{l=1}^{N}\sum_{m=1}^{N}\sum_{n=1}^{N} \left [ \underbrace{\left ( {\bf s}_\text{E}^l  \right )^\text{T} {\bf D}^{-1} {\bf c}_\text{E}^m}_{g_{m,l}}  \left ( {\bm{\delta}}^m \right )^\text{T} {\bf B}^m \left ( {\bf B}^n \right )^\text{T}{\bm{\delta}}^n \underbrace{\left ( {\bf c}_\text{E}^n  \right )^\text{T} {\bf D}^{-1} {\bf s}_\text{E}^l}_{f_{l,n}}  \right ]   \\
=& \frac{1}{4{\mu^2}}\sum_{l=1}^{N}\sum_{m=1}^{N}\sum_{n=1}^{N} \left\{ \left ( {\bm{ \delta}}^m \right )^\text{T} \left [ {p_{l,n}} {p_{m,l}} {\bf A}^m \left ( {\bf A}^n \right )^\text{T} \right ] {\bm{\delta}}^n  \right\} + \frac{1}{4{\mu^2}}\sum_{l=1}^{N}\sum_{m=1}^{N}\sum_{n=1}^{N} \left\{ \left ( {\bm{\delta}}^m \right )^\text{T} \left [ {f_{l,n}} {p_{m,l}} {\bf A}^m \left ( {\bf B}^n \right )^\text{T} \right ] {\bm{\delta}}^n  \right\} \\
& + \frac{1}{4{\mu^2}}\sum_{l=1}^{N}\sum_{m=1}^{N}\sum_{n=1}^{N} \left\{ \left ( {\bm{ \delta}}^m \right )^\text{T} \left [ {p_{l,n}} {g_{m,l}} {\bf B}^m \left ( {\bf A}^n \right )^\text{T} \right ] {\bm{\delta}}^n  \right\} + \frac{1}{4{\mu^2}}\sum_{l=1}^{N}\sum_{m=1}^{N}\sum_{n=1}^{N} \left\{ \left ( {\bm{\delta}}^m \right )^\text{T} \left [ {f_{l,n}} {g_{m,l}} {\bf B}^m \left ( {\bf B}^n \right )^\text{T} \right ] {\bm{\delta}}^n  \right\} \\
=& \frac{1}{4{\mu^2}}\sum_{l=1}^{N}\sum_{m=1}^{N}\sum_{n=1}^{N} \left\{ \left ( {\bm{ \delta}}^m \right )^\text{T} \left [ {p_{l,n}} {p_{m,l}} {\bf A}^m \left ( {\bf A}^n \right )^\text{T} + {f_{l,n}} {p_{m,l}} {\bf A}^m \left ( {\bf B}^n \right )^\text{T} + {p_{l,n}} {g_{m,l}} {\bf B}^m \left ( {\bf A}^n \right )^\text{T} + {f_{l,n}} {g_{m,l}} {\bf B}^m \left ( {\bf B}^n \right )^\text{T} \right ] {\bm{\delta}}^n  \right\}
\end{aligned}
\label{eq_29}
\end{equation}

\hrulefill
\vspace*{4pt}
\end{figure*}

\subsection{Dual Problem Formulation}
For the convex CI-BLP optimization problem $\mathcal{P}_1^\text{PSK}$ in \eqref{eq_13}, it is easy to validate that Slater's condition is satisfied, meaning that the dual gap is zero \cite{r35}. Thus, we can solve $\mathcal{P}_1^\text{PSK}$ by solving its dual problem, given by
\begin{equation}
\setcounter{equation}{21}
{\cal U}_1= \max_{\left\{{{\bm{\delta}}^m}\right\}, {\mu}} \min_{{\bf \hat W}, {t}}  {\cal L}_1\left ( {\bf \hat W} , {t}, {{\bm{\delta}}^m} , {\mu}  \right ),
\label{eq_21}
\end{equation}
where the inner minimization is achieved with \eqref{eq_16a}, the active power constraint in (17) and the expression for ${\bf \hat W}$ in \eqref{eq_20}. By substituting \eqref{eq_16a}, (17) and \eqref{eq_20} into ${\cal U}_1$ in (21), ${\cal U}_1$ can be further transformed and is shown in \eqref{eq_22} on the top of this page. By defining
\begin{equation}
\begin{aligned}
\setcounter{equation}{23}
p_{m,n}&=\left ( {{\bf s}_\text{E}^n} \right )^\text{T}{\bf C}^{-1} {{\bf s}_\text{E}^m}, {\kern 3pt} q_{m,n}= \left ( {{\bf c}_\text{E}^n} \right )^\text{T}{\bf C}^{-1} {{\bf c}_\text{E}^m}, \\
f_{m,n}&=\left ( {{\bf c}_\text{E}^n} \right )^\text{T}{\bf C}^{-1} {{\bf s}_\text{E}^m}, {\kern 3pt} g_{m,n}= \left ( {{\bf s}_\text{E}^n} \right )^\text{T}{\bf C}^{-1} {{\bf c}_\text{E}^m},
\end{aligned}
\label{eq_23}
\end{equation}
the objective function of the dual problem ${\cal U}_1$ can be further simplified and is given by \eqref{eq_24} above. Further defining
\begin{equation}
\setcounter{equation}{25}
{{\bm{\delta}}_\text{E}} = \left[ {\left ( {{\bm{\delta}}^1} \right ) ^\text{T}, \left ( {{\bm{\delta}}^2} \right ) ^\text{T}, \cdots, \left ( {{\bm{\delta}}^N} \right ) ^\text{T}} \right]^\text{T} \in {\mathbb R}^{2NK}
\label{eq_25}
\end{equation}
and ${\bf U}_{m,n} \in {\mathbb R}^{2K \times 2K}$ given by
\begin{equation}
\begin{aligned}
{\bf U}_{m,n}=&{\kern 3pt} p_{m,n} {\bf A}^m \left ( {\bf A}^n \right )^\text{T} + f_{m,n} {\bf A}^m \left ( {\bf B}^n \right )^\text{T} + g_{m,n} {\bf B}^m \left ( {\bf A}^n \right )^\text{T} \\
&+ q_{m,n} {\bf B}^m \left ( {\bf B}^n \right )^\text{T},
\end{aligned}
\label{eq_26}
\end{equation}
the objective function of the dual problem ${\cal U}_1$ can finally be obtained as
\begin{equation}
\begin{aligned}
{\cal U}_1 &=\min_{\left\{{{\bm{\delta}}^m}\right\}, {\mu}} \frac{1}{2 \mu} \sum_{m=1}^{N} \sum_{n=1}^{N} \left ( {{\bm{\delta}}^m} \right ) ^\text{T} {\bf U}_{m,n} {\bm{\delta}}^n \\
& =\min_{\left\{{{\bm{\delta}}^m}\right\}, {\mu}} \frac{1}{2 \mu} \left( {{{\bm{\delta}}_\text{E}}} \right)^\text{T} {\bf U} {{\bm{\delta}}_\text{E}},
\end{aligned}
\label{eq_27}
\end{equation}
where ${\bf U} \in {\mathbb R}^{2NK \times 2NK}$ is a block matrix constructed as
\begin{equation}
{\bf U} = \begin{bmatrix}
 {\bf U}_{1,1} & \cdots & \cdots& \cdots & {\bf U}_{1,N}\\
 \vdots & \ddots & {\bf U}_{m,n} & \ddots & \vdots \\
 {\bf U}_{N,1} & \cdots & \cdots & \cdots & {\bf U}_{N,N}
\end{bmatrix}.
\label{eq_28}
\end{equation}
The result in \eqref{eq_27} reveals that the objective function of the dual problem for the proposed CI-BLP scheme with PSK modulation is in a quadratic form.

After studying the objective function ${\cal U}_1$, we focus on the block-level power constraint in (17). By substituting the expression for $\bf \hat W$ in \eqref{eq_20} into (17), the transformation and simplification of the first term on the left-hand side of (17) is shown in \eqref{eq_29} on the top of this page. Based on the result in \eqref{eq_29} and by introducing ${\bf F}_{m,n}^l \in {\mathbb R}^{2K \times 2K}$ as
\begin{equation}
\begin{aligned}
\setcounter{equation}{30}
{\bf F}_{m,n}^l =& {\kern 3pt} {p_{l,n}} {p_{m,l}} {\bf A}^m \left ( {\bf A}^n \right )^\text{T} + {f_{l,n}} {p_{m,l}} {\bf A}^m \left ( {\bf B}^n \right )^\text{T} \\
&+ {p_{l,n}} {g_{m,l}} {\bf B}^m \left ( {\bf A}^n \right )^\text{T} + {f_{l,n}} {g_{m,l}} {\bf B}^m \left ( {\bf B}^n \right )^\text{T},
\label{eq_30}
\end{aligned}
\end{equation}
the first term on the left-hand side of (17) can finally be expressed in a compact form as
\begin{equation}
\begin{aligned}
\sum_{l=1}^{N} \left ( {\bf s}_\text{E}^l  \right )^\text{T} {\bf \hat W}^\text{T} {\bf \hat W} {\bf s}_\text{E}^l &= \frac{1}{4{\mu^2}}\sum_{l=1}^{N} \left( {{{\bm{\delta}}_\text{E}}} \right)^\text{T} {\bf F}^l {{\bm{\delta}}_\text{E}}\\
&= \frac{1}{4{\mu^2}} \left( {{{\bm{\delta}}_\text{E}}} \right)^\text{T} {\bf F}  {{\bm{ \delta}}_\text{E}},
\end{aligned}
\label{eq_31}
\end{equation}
where ${\bf F}= \sum_{l=1}^{N} {\bf F}^l \in {\mathbb R}^{2NK \times 2NK}$, with each ${\bf F}^l$ constructed as below:
\begin{equation}
{\bf F}^l = \begin{bmatrix}
 {\bf F}_{1,1}^l & \cdots & \cdots& \cdots & {\bf F}_{1,N}^l\\
 \vdots & \ddots & {\bf F}_{m,n}^l & \ddots & \vdots \\
 {\bf F}_{N,1}^l & \cdots & \cdots & \cdots & {\bf F}_{N,N}^l
\end{bmatrix}.
\label{eq_32}
\end{equation}
Following the above procedure, the second term on the left-hand side of (17) can be similarly expressed in a compact form as:
\begin{equation}
\sum_{n=1}^{N} \left ( {\bf c}_\text{E}^n  \right )^\text{T} {\bf \hat W}^\text{T} {\bf \hat W} {\bf c}_\text{E}^n = \frac{1}{4{\mu^2}} \left( {{{\bm{\delta}}_\text{E}}} \right)^\text{T} {\bf G}  {{\bm{ \delta}}_\text{E}},
\label{eq_33}
\end{equation}
where ${\bf G}= \sum_{l=1}^{N} {\bf G}^l$. Each ${\bf G}^l \in {\mathbb R}^{2NK \times 2NK}$, $\forall l \in {\cal N}$ is formulated similarly to \eqref{eq_32}, with each ${\bf G}_{m,n}^l \in {\mathbb R}^{2K \times 2K}$ given by
\begin{equation}
\begin{aligned}
{\bf G}_{m,n}^l = & {\kern 3pt} {g_{l,n}} {f_{m,l}} {\bf A}^m \left ( {\bf A}^n \right )^\text{T} + {q_{l,n}} {f_{m,l}} {\bf A}^m \left ( {\bf B}^n \right )^\text{T} \\
&+ {g_{l,n}} {q_{m,l}} {\bf B}^m \left ( {\bf A}^n \right )^\text{T} + {q_{l,n}} {q_{m,l}} {\bf B}^m \left ( {\bf B}^n \right )^\text{T}.
\end{aligned}
\label{eq_34}
\end{equation}
Based on the above derivations, the block-level power constraint (17) is equivalent to:
\begin{equation}
\begin{aligned}
&\frac{1}{4{\mu^2}} \left( {{{\bm{\delta}}_\text{E}}} \right)^\text{T} {\bf F}  {{\bm{ \delta}}_\text{E}} + \frac{1}{4{\mu^2}} \left( {{{\bm{\delta}}_\text{E}}} \right)^\text{T} {\bf G}  {{\bm{\delta}}_\text{E}} = Np_0 \\
\Rightarrow & \frac{1}{4{\mu^2}} \left( {{{\bm{\delta}}_\text{E}}} \right)^\text{T} \left({{\bf F}+{\bf G}}\right)  {{\bm{\delta}}_\text{E}} = Np_0.
\label{eq_35}
\end{aligned}
\end{equation}
Based on \eqref{eq_16a} and \eqref{eq_35}, we arrive at the following dual problem that optimizes ${\bm {\delta}}_\text{E}$ and $\mu$:
\begin{equation}
\begin{aligned}
&\mathcal{P}_2^\text{PSK}: {\kern 3pt} \min_{{\bm{\delta}}_\text{E}, {\mu}} \frac{1}{2{\mu}} {\bm {\delta}}_\text{E}^\text{T} {\bf U} {\bm {\delta}}_\text{E} \\
&{\kern 6pt} \text{s.t.} {\kern 11pt} {\bf C1:}{\kern 3pt} {\bf 1}^\text{T} {\bm {\delta}}_\text{E} -1=0 , \\
&{\kern 28.7pt} {\bf C2:}{\kern 3pt} \delta_\text{E}^m \ge 0, {\kern 3pt} \forall m \in \left\{ {1,2,\cdots, 2NK} \right\}, \\
&{\kern 29pt} {\bf C3:}{\kern 3pt} \frac{1}{4{\mu^2}} \left( {{{\bm{\delta}}_\text{E}}} \right)^\text{T} \left({{\bf F}+{\bf G}}\right)  {{\bm{\delta}}_\text{E}} = Np_0.
\label{eq_36}
\end{aligned}
\end{equation}

At first glance, $\mathcal{P}_2^\text{PSK}$ seems difficult to handle because $\mu$ is present in the denominator of the criterion as well as the equality power constraint. Nevertheless, we show below that after some transformations, $\mathcal{P}_2^\text{PSK}$ is equivalent to a QP optimization problem over a simplex.

We begin by studying the relationship between $\left( {{\bf F} + {\bf G}} \right)$ and $\bf U$ to further simplify $\mathcal{P}_2^\text{PSK}$, where the following proposition is obtained.

{\bf Proposition 1:} $\bf F$, $\bf G$, and $\bf U$ satisfy the following condition:
\begin{equation}
{\bf F} + {\bf G} = {\bf U}.
\label{eq_37}
\end{equation}

{\bf Proof:} See Appendix B.

According to {\bf Proposition 1} and based on the block-level power constraint in \eqref{eq_35}, we can obtain the following expression for $\mu$:
\begin{equation}
\frac{1}{4{\mu^2}} \left( {{{\bm{\delta}}_\text{E}}} \right)^\text{T} {\bf U}  {{\bm{\delta}}_\text{E}} = Np_0 {\kern 3pt}
\Rightarrow {\kern 3pt}  {\mu} = \sqrt{\frac{\left( {{{\bm{\delta}}_\text{E}}} \right)^\text{T} {\bf U}  {{\bm{\delta}}_\text{E}}}{4Np_0}}.
\label{eq_38}
\end{equation}
Substituting the above expression for $u_0$ into the objective function of the dual problem, ${\cal U}_1$ can be transformed into an optimization on ${{\bm{\delta}}_\text{E}}$ only, given by
\begin{equation}
\begin{aligned}
{\cal U}_1 &= \min_{{\bm{\delta}}_\text{E}, {\mu}} \frac{1}{2{\mu}} {\bm {\delta}}_\text{E}^\text{T} {\bf U} {\bm {\delta}}_\text{E}\\
& = \min_{{\bm{\delta}}_\text{E}} \frac{1}{2\sqrt{\frac{\left( {{{\bm{\delta}}_\text{E}}} \right)^\text{T} {\bf U}  {{\bm{\delta}}_\text{E}}}{4Np_0}}} {\bm {\delta}}_\text{E}^\text{T} {\bf U} {\bm {\delta}}_\text{E}\\
& = \min_{{\bm{\delta}}_\text{E}} \sqrt{N p_0 {\bm {\delta}}_\text{E}^\text{T} {\bf U} {\bm { \delta}}_\text{E}} \\
& = \min_{{\bm{\delta}}_\text{E}} {\bm {\delta}}_\text{E}^\text{T} {\bf U} {\bm {\delta}}_\text{E},
\label{eq_39}
\end{aligned}
\end{equation}
where the last step is achieved because $y=\sqrt{x}$ is a monotonic function. Accordingly, the final dual problem of the proposed CI-BLP optimization for PSK modulation can be formulated as
\begin{equation}
\begin{aligned}
&\mathcal{P}_3^\text{PSK}: {\kern 3pt} \min_{{\bm{\delta}}_\text{E}} {\bm {\delta}}_\text{E}^\text{T} {\bf U} {\bm {\delta}}_\text{E} \\
&{\kern 6pt} \text{s.t.} {\kern 11pt} {\bf C1:}{\kern 3pt} {\bf 1}^\text{T} {\bm {\delta}}_\text{E} -1=0 , \\
&{\kern 29pt} {\bf C2:}{\kern 3pt} \delta_\text{E}^m \ge 0, {\kern 3pt} \forall m \in \left\{ {1,2,\cdots, 2NK} \right\},
\label{eq_40}
\end{aligned}
\end{equation}
where we note that the block-level power constraint is no longer required because it is already inherently satisfied.

Comparing the QP formulation $\mathcal{P}_3^\text{PSK}$ for the proposed CI-BLP with that for traditional CI-SLP in the case of PSK modulation (shown as ${\cal P}_8$ in (50) of \cite{r31}), we observe that they share a similar problem formulation, with the only difference lying in the problem size. For traditional CI-SLP in \cite{r31}, the number of variables to be optimized is $2K$, while for the proposed CI-BLP it is $2NK$. This is because the proposed CI-BLP finds a single precoding matrix for all symbol slots, and thus the data symbols for the entire block need to be jointly considered. Nevertheless, we note that for the proposed CI-BLP scheme, the optimization only needs to be performed once per block, while the traditional CI-SLP needs to solve $N$ optimization problems for the considered block with length $N$, because traditional CI-SLP performs the optimization on a symbol-by-symbol basis. This will lead to a significant complexity reduction for the proposed CI-BLP over traditional CI-SLP, as will be shown in Section V.

$\mathcal{P}_3^\text{PSK}$ is a QP optimization problem over a simplex, which can be more efficiently solved than the original CI-BLP optimization problem $\mathcal{P}_1^\text{PSK}$ via the standard simplex method \cite{r36}, \cite{r37} or the interior-point methods \cite{r38}, which are not discussed in this paper for brevity. It is also worth mentioning that one advantage of the original problem formulation $\mathcal{P}_1^\text{PSK}$ over $\mathcal{P}_3^\text{PSK}$ is that the number of parameters to be optimized does not grow with the size $N$ of the block. After solving $\mathcal{P}_3^\text{PSK}$ and obtaining ${\bf \hat W}$ via \eqref{eq_20}, the original complex precoding matrix $\bf W$ in \eqref{eq_1} for PSK modulation can be obtained by
\begin{equation}
{\bf W}= {\bf \hat W} {\bf \hat P} - \jmath {\bf \hat W} {\bf \hat  Q},
\label{eq_41}
\end{equation}
where the form of ${\bf \hat P}$ and ${\bf \hat Q}$ follows \eqref{eq_9} while their dimension is changed into $2K \times K$.

\begin{figure}[!b]
\centering
\includegraphics[scale=0.65]{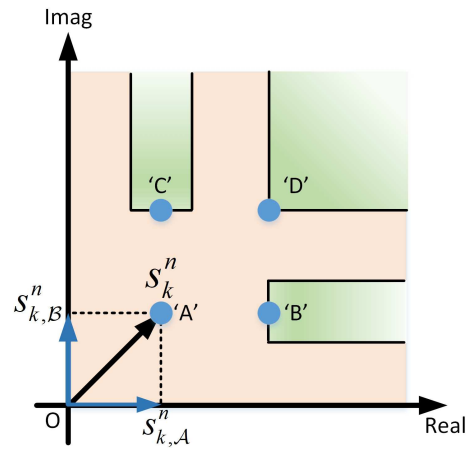}
\caption{An illustration for CI, 16QAM, `symbol-scaling' metric}
\end{figure}

\newcounter{mytempeqncnt7}
\begin{figure*}[!t]
\normalsize
\setcounter{equation}{44}

\begin{equation}
\begin{aligned}
&{\cal L}_2\left ( {\bf \hat W} , {\hat t}, {\hat \delta_{k}^{n}}, {\hat \vartheta_{k}^{n}}, {\hat \mu}  \right ) \\
& = -{\hat t} +\sum_{n=1}^{N} \sum_{k \in \text{card}\left\{{\cal O}^n\right\}} {\hat \delta_k^n}\left [ {{\hat t}- \left( {{\bf a}_k^n} \right)^\text{T}{\bf \hat W} {\bf s}_\text{E}^n - \left( {{\bf b}_k^n} \right)^\text{T}{\bf \hat W} {\bf c}_\text{E}^n} \right ] +\sum_{n=1}^{N} \sum_{k \in \text{card}\left\{{\cal I}^n\right\}} {\hat \vartheta_k^n}\left [ {{\hat t}- \left( {{\bf a}_k^n} \right)^\text{T}{\bf \hat W} {\bf s}_\text{E}^n - \left( {{\bf b}_k^n} \right)^\text{T}{\bf \hat W} {\bf c}_\text{E}^n} \right ] \\
& {\kern 10pt} + {\hat \mu} \left[{\sum_{n=1}^{N} \left ( {\bf s}_\text{E}^n \right )^\text{T} \left ( {{\bf P} {\bf \hat W} + {\bf Q} {\bf \hat W} {\bf T}} \right )^\text{T}\left ( {{\bf P} {\bf \hat W} + {\bf Q} {\bf \hat W} {\bf T}} \right ) {\bf s}_\text{E}^n -N p_0 }\right] \\
&=\left ( {\sum_{n=1}^{N} {\bf 1}^\text{T}{\bm{\hat \delta}}^n + \sum_{n=1}^{N} {\bf 1}^\text{T}{\bm{\hat \vartheta}}^n -1 } \right ) {\hat t} -\sum_{n=1}^{N}\left( {{\bm{\hat \delta}}^n} \right)^\text{T}{\bf A}_{\cal O}^n{\bf \hat W}{\bf s}_\text{E}^n -\sum_{n=1}^{N}\left( {{\bm{\hat \delta}}^n} \right)^\text{T}{\bf B}_{\cal O}^n{\bf \hat W}{\bf c}_\text{E}^n -\sum_{n=1}^{N}\left( {{\bm{\hat \vartheta}}^n} \right)^\text{T}{\bf A}_{\cal I}^n{\bf \hat W}{\bf s}_\text{E}^n \\
& {\kern 10pt} -\sum_{n=1}^{N}\left( {{\bm{\hat \vartheta}}^n} \right)^\text{T}{\bf B}_{\cal I}^n{\bf \hat W}{\bf c}_\text{E}^n + {\hat \mu} \left[{\sum_{n=1}^{N} \left ( {\bf s}_\text{E}^n \right )^\text{T} \left ( { {\bf \hat W}^\text{T} {\bf P}^\text{T}  + {\bf T}^\text{T} {\bf \hat W}^\text{T}  {\bf Q}^\text{T}  } \right )\left ( {{\bf P} {\bf \hat W} + {\bf Q} {\bf \hat W} {\bf T}} \right ) {\bf s}_\text{E}^n }\right] - {\hat \mu}N p_0\\
&=\left ( {\sum_{n=1}^{N} {\bf 1}^\text{T}{\bm{\beta}}^n -1 } \right ) {\hat t} -\sum_{n=1}^{N}\left( {{\bm{ \beta}}^n} \right)^\text{T}{\bf A}^n{\bf \hat W}{\bf s}_\text{E}^n -\sum_{n=1}^{N}\left( {{\bm{\beta}}^n} \right)^\text{T}{\bf B}^n{\bf \hat W}{\bf c}_\text{E}^n + {\hat \mu}\sum_{n=1}^{N} \left ( {\bf s}_\text{E}^n  \right )^\text{T} {\bf \hat W}^\text{T} {\bf P}^\text{T}{\bf P}{\bf \hat W} {\bf s}_\text{E}^n  \\
& {\kern 10pt} + \underbrace{{\hat \mu}\sum_{n=1}^{N} \left ( {\bf s}_\text{E}^n  \right )^\text{T} {\bf \hat W}^\text{T} {\bf P}^\text{T}{\bf Q}{\bf \hat W}{\bf T}{\bf s}_\text{E}^n}_{=0}  + \underbrace{{\hat \mu}\sum_{n=1}^{N} \left ( {\bf s}_\text{E}^n  \right )^\text{T} {\bf T}^\text{T} {\bf \hat W}^\text{T}  {\bf Q}^\text{T} {\bf P} {\bf \hat W} {\bf s}_\text{E}^n}_{=0}  + {\hat \mu}\sum_{n=1}^{N} \underbrace{\left ( {\bf s}_\text{E}^n  \right )^\text{T} {\bf T}^\text{T}}_{\left( {\bf c}_\text{E}^n  \right)^\text{T} }  {\bf \hat W}^\text{T}  {\bf Q}^\text{T}{\bf Q}{\bf \hat W}\underbrace{{\bf T}{\bf s}_\text{E}^n}_{{\bf c}_\text{E}^n}  -{\hat \mu} N p_0 \\
&=\left ( {\sum_{n=1}^{N} {\bf 1}^\text{T}{\bm{\beta}}^n -1 } \right ) {\hat t} -\sum_{n=1}^{N}\left( {{\bm{ \beta}}^n} \right)^\text{T}{\bf A}^n{\bf \hat W}{\bf s}_\text{E}^n -\sum_{n=1}^{N}\left( {{\bm{\beta}}^n} \right)^\text{T}{\bf B}^n{\bf \hat W}{\bf c}_\text{E}^n + {\hat \mu}\sum_{n=1}^{N} \left ( {\bf s}_\text{E}^n  \right )^\text{T} {\bf \hat W}^\text{T} {\bf \hat W} {\bf s}_\text{E}^n + {\hat \mu}\sum_{n=1}^{N} \left ( {\bf c}_\text{E}^n  \right )^\text{T} {\bf \hat W}^\text{T} {\bf \hat W} {\bf c}_\text{E}^n -{\hat \mu}Np_0
\end{aligned}
\label{eq_45}
\end{equation}

\hrulefill
\vspace*{4pt}
\end{figure*}

\section{Proposed CI-BLP for QAM Modulation}
\subsection{Problem Formulation}
In this section, we extend the proposed CI-BLP approach to the case of QAM modulation, where we still employ the `symbol-scaling' CI metric. Compared with PSK, one significant difference for QAM modulation is that not all QAM constellation points can exploit CI. To be more specific, the constellation points for QAM can be divided into 4 types, as shown in Fig. 3 where one quarter of a nominal 16QAM constellation is depicted as a representative example:

\begin{enumerate}
  \item Type `A': no CI can be exploited;
  \item Type `B': CI can be exploited for the real part;
  \item Type `C': CI can be exploited for the imaginary part;
  \item Type `D': CI can be exploited for both the real and imaginary part.
\end{enumerate}

By following a similar procedure as that in Section III, we decompose each data symbol $s_k^n$ and its corresponding received signal excluding noise as in (2), where for QAM modulation we have
\begin{equation}
\setcounter{equation}{42}
s_{k,{\cal A}}^n=\Re \left( s_k^n \right), {\kern 3pt} s_{k,{\cal B}}^n=\jmath \Im \left( s_k^n \right), {\kern 3pt} \forall k,n.
\label{eq_42}
\end{equation}
Introducing ${\bm \alpha}_\text{E}^n$ as in \eqref{eq_3} and following a similar formulation as in \eqref{eq_4}-\eqref{eq_5} and \eqref{eq_7}-\eqref{eq_12}, the proposed CI-BLP optimization problem for QAM modulation can be formulated as:
\begin{equation}
\begin{aligned}
&\mathcal{P}_1^\text{QAM}: {\kern 3pt} \min_{{\bf \hat W}, {\hat t}} - {\hat t} \\
&\text{s.t.} {\bf C1:}{\kern 2pt} {\hat t} - \left( {{\bf a}_k^n} \right)^\text{T}{\bf \hat W} {\bf s}_\text{E}^n - \left( {{\bf b}_k^n} \right)^\text{T}{\bf \hat W} {\bf c}_\text{E}^n \le 0, {\kern 1pt} \forall k \in {\cal O}^n, n\in {\cal N}, \\
&{\kern 12pt} {\bf C2:}{\kern 2pt} {\hat t} - \left( {{\bf a}_k^n} \right)^\text{T}{\bf \hat W} {\bf s}_\text{E}^n - \left( {{\bf b}_k^n} \right)^\text{T}{\bf \hat W} {\bf c}_\text{E}^n = 0, {\kern 1pt} \forall k \in {\cal I}^n, n\in {\cal N}, \\
&{\kern 12pt} {\bf C3:}{\kern 2pt} \sum_{n=1}^{N} \left \| {\left({{\bf P} {\bf \hat W} + {\bf Q} {\bf \hat W} {\bf T}}\right) {\bf s}_\text{E}^n  }  \right \| _2^2 - N{p_0} \le 0.
\label{eq_43}
\end{aligned}
\end{equation}
Recalling ${\cal K}=\left\{ {1,2,\cdots,2K} \right\}$, the set ${\cal O}^n$ consists of the indices in $\cal K$ that correspond to the real part of the data symbol belonging to constellation point type `B', the imaginary part of the data symbol belonging to constellation point type `C', and both the real and imaginary part of the data symbol belonging to constellation point type `D', for which CI can be exploited. The set ${\cal I}^n$ consists of the indices in $\cal K$ that correspond to the real part of the data symbol belonging to constellation point type `C', the imaginary part of the data symbol belonging to constellation point type `B', and both the real and imaginary part of the data symbol belonging to constellation point type `A', for which there exists no CI to be exploited. ${\cal O}^n$ and ${\cal I}^n$ satisfy:
\begin{equation}
\begin{aligned}
&{\kern 18pt} {\cal O}^n \cup {\cal I}^n = {\cal K}, {\kern 3pt} {\cal O}^n \cap {\cal I}^n = \varnothing, \\
&\text{card}\left\{ {\cal O}^n \right\}+\text{card}\left\{ {\cal I}^n \right\}=2K, {\kern 3pt} \forall n \in {\cal N}.
\end{aligned}
\label{eq_44}
\end{equation}

Similar to the case for PSK, the proposed CI-BLP scheme for QAM offers complexity reduction over traditional CI-SLP, because only a single optimization problem needs to be solved per transmission block. More importantly, another advantage that is specific to QAM modulation over traditional CI-SLP is the reduction in signaling overhead. Since traditional CI-SLP is performed on a symbol level, the power normalization factor varies from symbol to symbol. Therefore, extra signaling overhead is required for CI-SLP since the power normalization factor needs to be broadcast to the users at the symbol rate for correct demodulation, while the proposed CI-BLP scheme returns a constant power normalization factor (equal to the optimal value $t^*$ for ${\cal P}_1^\text{QAM}$) over the entire block, and thus it enjoys reduced signaling overhead compared with CI-SLP.

\newcounter{mytempeqncnt8}
\begin{figure*}[!t]
\normalsize
\setcounter{equation}{46}

\begin{IEEEeqnarray}{rCl}
\IEEEyesnumber
\frac{{\partial {\cal L}_2}}{{\partial \hat t}} =\sum_{n=1}^{N} {\bf 1}^\text{T}{\bm{\beta}}^n -1 =0  {\kern 30pt} \IEEEyessubnumber* \label{eq_47a} \\
\frac{{\partial {\cal L}_2}}{{\partial {\bf \hat W}}} = -\sum_{n=1}^{N} \left [ {\left ( {{\bm{\beta}}^n} \right ) }^\text{T} {\bf A}^n \right ] ^\text{T} \left ( {\bf s}_\text{E}^n  \right )^\text{T} -\sum_{n=1}^{N} \left [ {\left ( {{\bm{\beta}}^n} \right ) }^\text{T} {\bf B}^n \right ] ^\text{T} \left ( {\bf c}_\text{E}^n  \right )^\text{T} + 2{\hat \mu} {\bf \hat W}\left [ {\sum_{n=1}^{N} {\bf s}_\text{E}^n  \left ( {\bf s}_\text{E}^n \right )^\text{T} + \sum_{n=1}^{N} {\bf c}_\text{E}^n  \left ( {\bf c}_\text{E}^n \right )^\text{T} }  \right ]  = {\bf 0}  {\kern 30pt} \label{eq_47b} \\
{\hat \delta_k^n}\left [ {{\hat t}- \left( {{\bf a}_k^n} \right)^\text{T}{\bf \hat W} {\bf s}_\text{E}^n - \left( {{\bf b}_k^n} \right)^\text{T}{\bf \hat W} {\bf c}_\text{E}^n} \right ]=0, {\kern 3pt} {\hat \delta_k^n} \ge 0, {\kern 3pt} \forall k \in {\cal O}^n, {\kern 3pt} n \in {\cal N} {\kern 30pt} \label{eq_47c} \\
{{\hat t}- \left( {{\bf a}_k^n} \right)^\text{T}{\bf \hat W} {\bf s}_\text{E}^n - \left( {{\bf b}_k^n} \right)^\text{T}{\bf \hat W} {\bf c}_\text{E}^n}=0, {\kern 3pt} \forall k \in {\cal I}^n, {\kern 3pt} n \in {\cal N} {\kern 30pt} \label{eq_47d} \\
{\hat \mu}\left [ \sum_{n=1}^{N} \left ( {\bf s}_\text{E}^n  \right )^\text{T} {\bf \hat W}^\text{T} {\bf \hat W} {\bf s}_\text{E}^n + \sum_{n=1}^{N} \left ( {\bf c}_\text{E}^n  \right )^\text{T} {\bf \hat W}^\text{T} {\bf \hat W} {\bf c}_\text{E}^n -Np_0 \right ] =0, {\kern 3pt} {\hat \mu} \ge 0 {\kern 30pt} \label{eq_47e}
\end{IEEEeqnarray}

\hrulefill
\vspace*{4pt}
\end{figure*}

\subsection{Optimal Closed-Form Structure for $\bf \hat W$}
Similar to the case when PSK modulation is considered, we derive the optimal precoding matrix $\bf \hat W$ for $\mathcal{P}_1^\text{QAM}$ based on the Lagrangian and KKT conditions. To begin with, the Lagrangian of $\mathcal{P}_1^\text{QAM}$ can be constructed and is shown in \eqref{eq_45} on the top of previous page, where ${\bm{\hat \delta}}^n \in {\mathbb R}^{\text{card}\left\{ {\cal O}^n \right\}}$ and ${\bm{\hat \vartheta }}^n \in {\mathbb R}^{\text{card}\left\{ {\cal I}^n \right\}}$ consist of the Lagrange multipliers that correspond to the inequality and equality constraints, respectively. For each symbol slot $n \in {\cal N}$, ${\bf A}_{\cal O}^n$ and ${\bf B}_{\cal O}^n$ consist of the rows of ${\bf A}^n$ and ${\bf B}^n$ respectively, whose indices belong to ${\cal O}^n$, while ${\bf A}_{\cal I}^n$ and ${\bf B}_{\cal I}^n$ consist of the rows of ${\bf A}^n$ and ${\bf B}^n$, whose indices belong to ${\cal I}^n$. ${\bm \beta}^n = \left[ {\beta_1^n,\beta_2^n,\cdots,\beta_K^n} \right] \in {\mathbb R}^{2K}$, with the $k$-th entry defined as:
\begin{equation}
\setcounter{equation}{46}
\beta_k^n= \left\{\begin{matrix}
\hat \delta_k^n, {\kern 3pt} \text{if} {\kern 3pt} k \in {\cal O}^n   \\
\hat \vartheta_k^n, {\kern 3pt} \text{if} {\kern 3pt} k \in {\cal I}^n
\end{matrix}\right. ,{\kern 3pt} \forall k \in {\cal K}.
\label{eq_46}
\end{equation}

The main difference between the Lagrangian function ${\cal L}_2$ for QAM and the Lagrangian function ${\cal L}_1$ for PSK lies in the fact that the CI-BLP optimization for QAM includes equality constraints, as shown in the 3rd constraint for $\mathcal{P}_1^\text{QAM}$ in \eqref{eq_43}. This observation means that not all the entries in ${\bm \beta}^n$ need to be non-negative, which will mean that the final QP optimization for QAM modulation is not over a simplex any more. This is evident from the KKT conditions of the Lagrangian function ${\cal L}_2$ shown in (47) on the top of this page. As can be observed in \eqref{eq_47c} and \eqref{eq_47d}, the value of each ${\hat \delta_k^n}$ has to be non-negative, while the value of each $\hat \vartheta_k^n$ does not because it corresponds to the equality constraint.

Similar to the case for PSK modulation in Section III, when the optimal solution to $\mathcal{P}_1^\text{QAM}$ is obtained, the expression for the precoding matrix $\bf \hat W$ as a function of the Lagrange multipliers is found from \eqref{eq_47b} to be
\begin{equation}
\setcounter{equation}{48}
{\bf \hat W} = \frac{1}{2{\hat \mu}} \sum_{n=1}^{N} \left [ {\left ( {\bf A}^n \right ) }^\text{T} {{\bm{\beta}}^n} \left ( {\bf s}_\text{E}^n  \right )^\text{T} + {\left ( {\bf B}^n \right ) }^\text{T} {{\bm{\beta}}^n} \left ( {\bf c}_\text{E}^n  \right )^\text{T} \right ] {\bf D}^{-1}.
\label{eq_48}
\end{equation}

\subsection{Dual Problem Formulation}
The procedure for deriving the dual problem formulation for QAM is similar to that outlined in Section III-C for PSK, and therefore for brevity we directly present the final QP optimization problem for QAM modulation below:
\begin{equation}
\begin{aligned}
&\mathcal{P}_2^\text{QAM}: {\kern 3pt} \min_{{\bm{\beta}}_\text{E}} {\bm {\beta}}_\text{E}^\text{T} {\bf U} {\bm {\beta}}_\text{E} \\
&{\kern 6pt} \text{s.t.} {\kern 11pt} {\bf C1:}{\kern 3pt} {\bf 1}^\text{T} {\bm {\beta}}_\text{E} -1=0 , \\
&{\kern 29pt} {\bf C2:}{\kern 3pt} \beta_\text{E}^m \ge 0, {\kern 3pt} \forall m \in {\cal M},
\label{eq_49}
\end{aligned}
\end{equation}
where ${\bm \beta}_\text{E} \in {\mathbb R}^{2NK}$ is defined as
\begin{equation}
{\bm \beta}_\text{E} = \left[ {{\bm \beta}^1, {\bm \beta}^2, \cdots, {\bm \beta}^N} \right]^\text{T}.
\label{eq_50}
\end{equation}
${\cal M}$ contains the indices of the Lagrange multipliers corresponding to the inequality constraints for all symbol slots within the considered block, and is expressed mathematically as
\begin{equation}
{\cal M}=\left \{ m \mid m=2\left ( n-1 \right )K +k, {\kern 3pt} \text{if} {\kern 3pt} \beta_k^n = \hat \delta_k^n  \right \}.
\label{eq_51}
\end{equation}
Compared to the QP formulation $\mathcal{P}_3^\text{PSK}$ for PSK in \eqref{eq_40}, the QP formulation $\mathcal{P}_2^\text{QAM}$ for QAM only requires a total number of $\text{card}\left\{ \cal M \right\}$ entries in ${\bm \beta}_\text{E}$ to be non-negative. In this case, although the simplex method cannot be used for solving $\mathcal{P}_2^\text{QAM}$, the interior-point based methods \cite{r38} can still be employed to efficiently solve $\mathcal{P}_2^\text{QAM}$. After obtaining ${\bm \beta}_\text{E}$, the precoding matrix $\bf \hat W$ can be obtained via (48), and the final complex precoding matrix $\bf W$ for QAM modulation can be obtained by \eqref{eq_41}.

\section{Numerical Results}
In this section, numerical results are presented based on Monte Carlo simulations. In each plot, we assume that the transmit power budget per symbol slot is $p_0=1$, leading to the total transmit power budget for the considered block of symbol slots as $P_\text{total}=N p_0 =N$. We compare our proposed CI-BLP schemes with closed-form ZF-based methods and traditional CI-SLP methods for both PSK and QAM modulation.

The following abbreviations are used throughout this section:
\begin{enumerate}
\item `ZF': Traditional ZF precoding with block-level power normalization, where the precoding matrix is given by
\begin{equation}
{{\bf{W}}_\text{ZF}} = \frac{1}{f_\text{ZF}} {{\bf{H}}^\text{H}}{\left( {{\bf{H}}{{\bf{H}}^\text{H}}} \right)^{ - 1}},
\label{eq_69}
\end{equation}
with block-level scaling factor
\begin{equation}
f_\text{ZF} = \sqrt{\frac{{{\bf{W}}_\text{ZF}} {\bf S}}{N p_0}} ;
\label{eq_70}
\end{equation}
\item `RZF': Traditional RZF precoding with block-level power normalization, where the precoding matrix is given by
\begin{equation}
{{\bf{W}}_\text{RZF}} = \frac{1}{f_\text{RZF}} {{\bf{H}}^\text{H}}{\left( {{\bf{H}}{{\bf{H}}^\text{H}} + \frac{K}{\rho } {\bf{I}}} \right)^{ - 1}},
\label{eq_71}
\end{equation}
with block-level scaling factor
\begin{equation}
f_\text{RZF} = \sqrt{\frac{{{\bf{W}}_\text{RZF}} {\bf S}}{N p_0}} ;
\label{eq_72}
\end{equation}
\item `CI-SLP': Traditional CI-SLP method with symbol-level power constraint for PSK in \cite{r31} and QAM in \cite{r32};
\item `CI-BLP-CVX': Original CI-BLP optimization ${\cal P}_{1}^\text{PSK}$ in \eqref{eq_13} and ${\cal P}_{1}^\text{QAM}$ in \eqref{eq_43};
\item `CI-BLP-QP': Proposed QP solution for CI-BLP ${\cal P}_{3}^\text{PSK}$ in \eqref{eq_40} and ${\cal P}_{2}^\text{QAM}$ in \eqref{eq_49}.
\end{enumerate}

\begin{figure}[!t]
\centering
\includegraphics[scale=0.45]{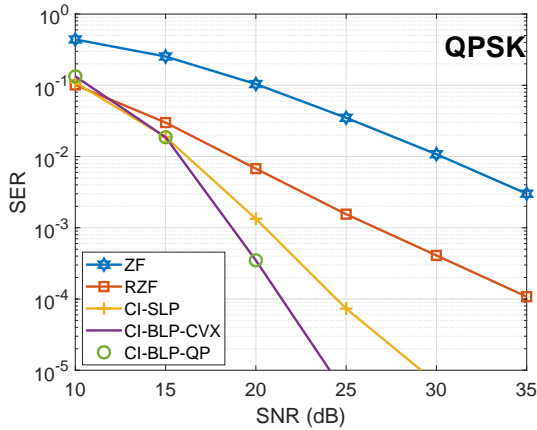}
\caption{SER v.s. SNR, QPSK, $N_\text{T}=K=12$, $N=15$}
\end{figure}

Fig. 4 depicts the symbol error rate (SER) of the proposed CI-BLP scheme when QPSK modulation is employed in a $12 \times 12$ MU-MISO system, where the length of the block is $N=15$. As can be observed, both CI-based precoding approaches achieve an improved performance over ZF precoding. When the length of the transmission block is short, we observe that the proposed CI-BLP offers noticeable performance gains over traditional CI-SLP that optimizes the precoding matrix on a symbol level, owing to the relaxed power constraint over the entire block, i.e., a power allocation among symbol slots is inherently performed for the proposed CI-BLP method. The result in Fig. 4 also validates the correctness of our derivations in the paper, as evidenced by the identical SER performance for `CI-BLP-CVX' and `CI-BLP-QP'.

\begin{figure}[!t]
\centering
\includegraphics[scale=0.45]{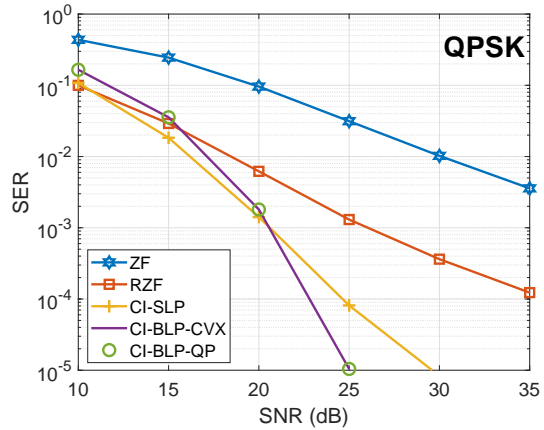}
\caption{SER v.s. SNR, QPSK, $N_\text{T}=K=12$, $N=40$}
\end{figure}

Fig. 5 compares the SER performance of different CI precoding approaches for a $12 \times 12$ MU-MISO system with QPSK modulation, where the length of the block is increased to $N=40$. In this case, the proposed CI-BLP still offers significant performance gains over conventional ZF and RZF precoding. While we observe that the performance gain of the proposed CI-BLP becomes less significant when the block length increases, it still outperforms traditional CI-SLP when the transmit SNR goes above 20dB. Compared to traditional CI-SLP methods that need to solve $N$ optimization problems for the block, the proposed CI-BLP method only needs to solve the CI-BLP optimization problem once, thus further motivating the use of CI-based precoding in practical wireless systems.

\begin{figure}[!t]
\centering
\includegraphics[scale=0.45]{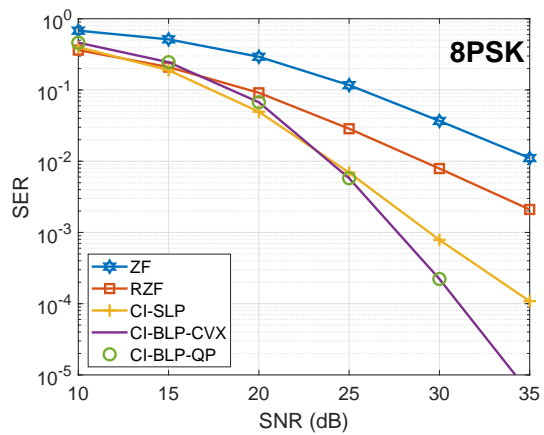}
\caption{SER v.s. SNR, 8PSK, $N_\text{T}=K=12$, $N=15$}
\end{figure}

Fig. 6 depicts the SER performance of different precoding methods when 8PSK modulation is employed, where $K=N_\text{T}=12$ and the block length is $N=15$. Similar to the QPSK case, both CI-based precoding approaches outperform conventional ZF and RZF precoding. Again, as the transmit SNR increases, the proposed CI-BLP is able to achieve an improved SER over the traditional CI-SLP scheme, thanks to the inherent power allocation among different symbol slots.

\begin{figure}[!t]
\centering
\includegraphics[scale=0.45]{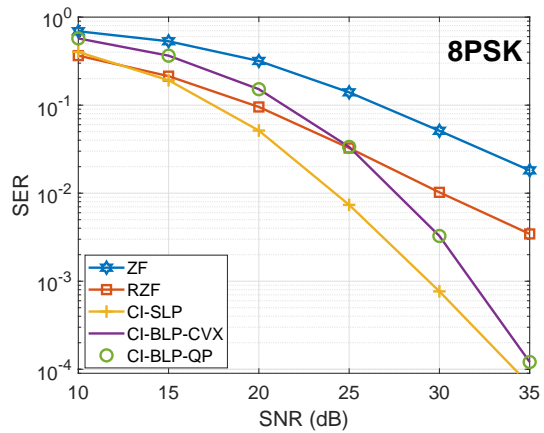}
\caption{SER v.s. SNR, 8PSK, $N_\text{T}=K=12$, $N=40$}
\end{figure}

Fig. 7 compares the SER performance of different precoding schemes when the length of the transmission block is increased to $N=40$ for 8PSK modulation,  where $K=N_\text{T}=12$. As $N$ increases, the performance gain of the proposed CI-BLP method is reduced due to that fact that a larger number of constraints are simultaneously enforced in the corresponding CI-BLP optimization problem, but it is still able to outperform traditional ZF and RZF precoding when the transmit SNR is larger than 25dB and approach the traditional CI-SLP scheme. Again, the numerical results validate the correctness of our derivations for PSK modulation in Section III.

\begin{figure}[!t]
\centering
\includegraphics[scale=0.45]{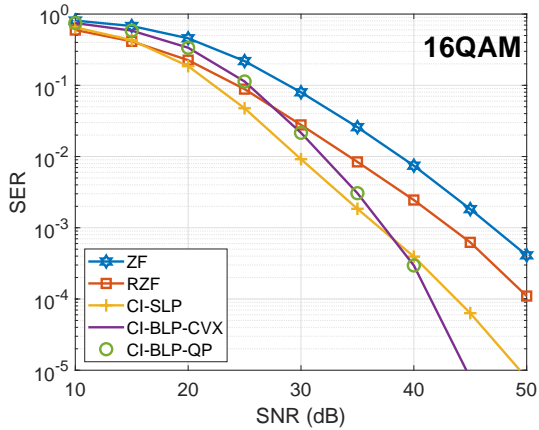}
\caption{SER v.s. SNR, 16QAM, $N_\text{T}=K=12$, $N=15$}
\end{figure}

Fig. 8 depicts the SER performance of different precoding approaches when QAM modulation is employed for a $12 \times 12$ MU-MISO communication system, with a block length of $N=15$. As can be observed, when we shift from PSK modulation to QAM modulation, the SER improvements for CI-based precoding become less significant, because only the outer QAM constellation points can exploit CI. The result in Fig. 8 validates the correctness of our derivations for the proposed CI-BLP with QAM modulation in Section IV, as evidence by the identical SER performance for `CI-BLP-CVX' and `CI-BLP-QP'.

\begin{figure}[!t]
\centering
\includegraphics[scale=0.46]{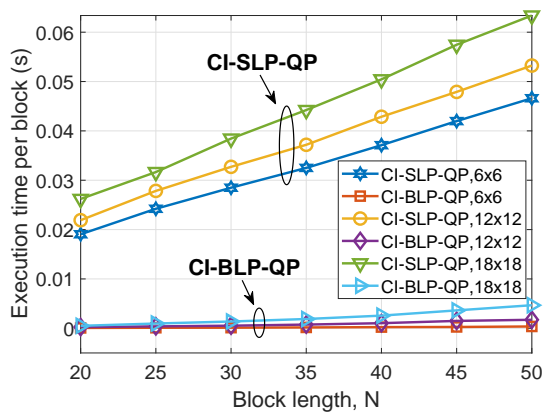}
\caption{Execution time v.s. block length $N$}
\end{figure}

Fig. 9 depicts the complexity of the proposed CI-BLP method with traditional CI-SLP in terms of the execution time running on a Windows 10 Laptop with i7-11390H and 16GB RAM, where results for $6 \times 6$, $12 \times 12$ and $18 \times 18$ MU-MISO systems are presented. For fairness of comparison, we only evaluate the time complexity of the `quadprog' function in MATLAB that is used to solve the corresponding QP problem for both CI-BLP and CI-SLP, and avoid the run time consumed in constructing the required matrices/vectors. Since the size of the QP problem is independent of the modulation type, the modulation does not significantly affect the complexity, which is determined primarily by the number of users and transmit antennas. From Fig. 9, we observe that the proposed CI-BLP approach offers a significant complexity gain over traditional CI-SLP, and the complexity gains become more prominent as the block length $N$ increases.

\begin{figure}[!t]
\centering
\includegraphics[scale=0.45]{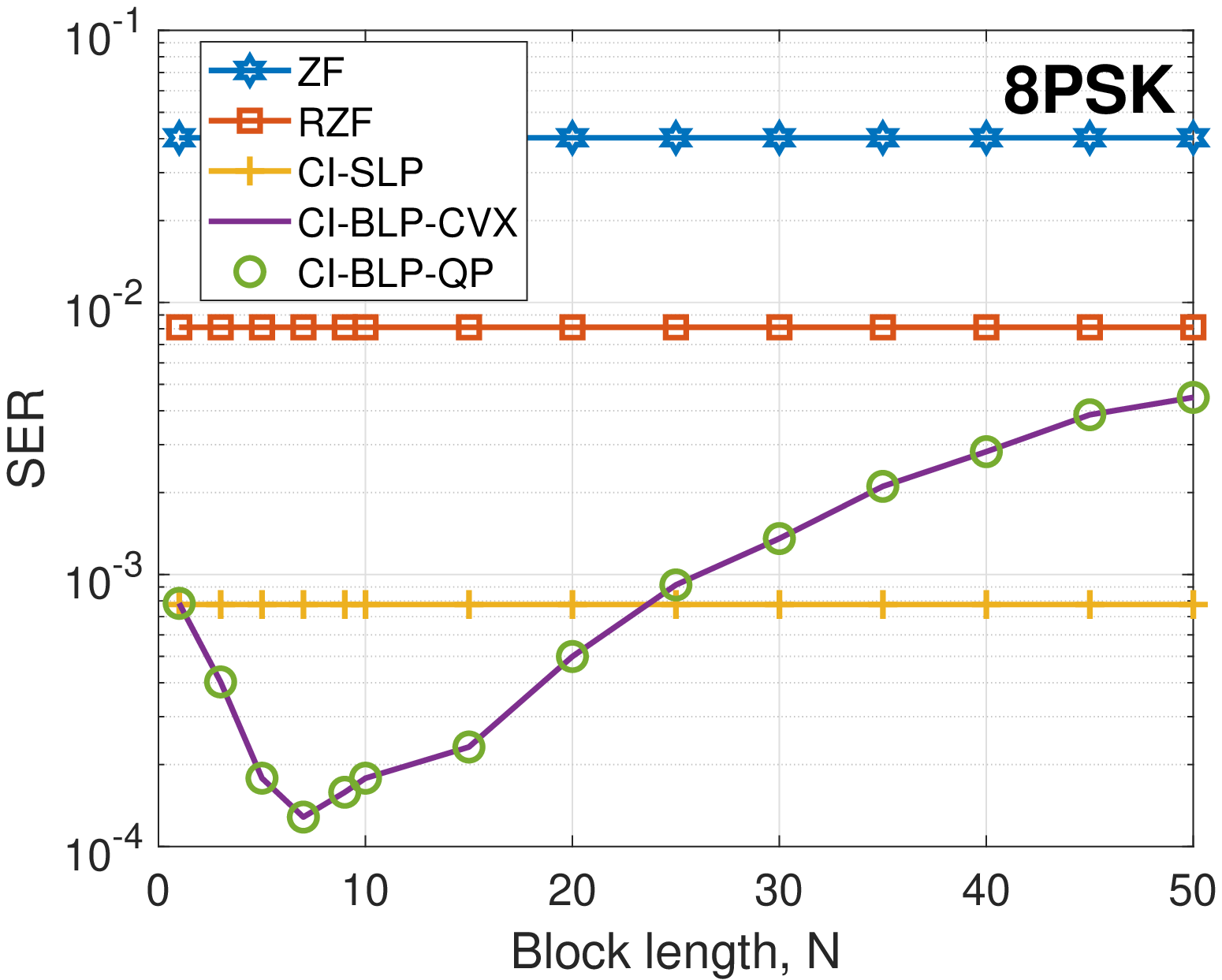}
\caption{SER v.s. block length $N$, 8PSK, $N_\text{T}=K=12$, $\text{SNR}=30\text{dB}$}
\end{figure}

To illustrate the effect of block length $N$ on the performance, in Fig. 10 we depict the SER with respect to the block length $N$, where 8PSK modulation is employed at a transmit SNR of 30dB. As can be observed, CI-BLP returns the same SER performance as CI-SLP when $N=1$, because CI-BLP reduces to CI-SLP when optimized independently for each symbol slot. Interestingly, as the block length $N$ increases, we observe that the SER performance firstly improves since the benefit of the relaxed power constraint outweighs the loss due to using a fixed precoder over the block, while the SER performance becomes worse as $N$ further increases, because the benefit of the relaxed power constraint cannot further compensate for the loss of the fixed precoder. The results in Fig. 9 and Fig. 10 demonstrates that CI-BLP achieves an improved performance-complexity tradeoff when the value of the block length $N$ is moderate.

\newcounter{mytempeqncnt6}
\begin{figure*}[!t]
\normalsize
\setcounter{equation}{58}

\begin{equation}
\begin{aligned}
&{\kern 3pt} {\bf F}_{m,n} + {\bf G}_{m,n} \\
= &{\kern 3pt} \sum_{l=1}^{N}{\bf F}_{m,n}^l + \sum_{l=1}^{N}{\bf G}_{m,n}^l \\
= &{\kern 2pt} \sum_{l=1}^{N} \left ( {p_{l,n}p_{m,l}+g_{l,n}f_{m,l}} \right ) {\bf A}^m \left ( {\bf A}^n  \right )^\text{T} + \sum_{l=1}^{N} \left ( {p_{l,n}p_{m,l}+g_{l,n}f_{m,l}} \right ) {\bf A}^m \left ( {\bf B}^n  \right )^\text{T} + \sum_{l=1}^{N} \left ( {p_{l,n}g_{m,l}+g_{l,n}q_{m,l}} \right ) {\bf B}^m \left ( {\bf A}^n  \right )^\text{T} \\
 &{\kern 2pt} + \sum_{l=1}^{N} \left ( {f_{l,n}g_{m,l}+q_{l,n}q_{m,l}} \right ) {\bf B}^m \left ( {\bf B}^n  \right )^\text{T} \\
= &{\kern 2pt} \sum_{l=1}^{N} \left [ \left ( {{\bf s}_\text{E}^n} \right )^\text{T}{\bf D}^{-1} {{\bf s}_\text{E}^l} \left ( {{\bf s}_\text{E}^l} \right )^\text{T}{\bf D}^{-1} {{\bf s}_\text{E}^m} + \left ( {{\bf s}_\text{E}^n} \right )^\text{T}{\bf D}^{-1} {{\bf c}_\text{E}^l} \left ( {{\bf c}_\text{E}^l} \right )^\text{T}{\bf D}^{-1} {{\bf s}_\text{E}^m} \right ] {\bf A}^m \left ( {\bf A}^n \right )^\text{T} \\
 &{\kern 2pt} + \sum_{l=1}^{N} \left [ \left ( {{\bf c}_\text{E}^n} \right )^\text{T}{\bf D}^{-1} {{\bf s}_\text{E}^l} \left ( {{\bf s}_\text{E}^l} \right )^\text{T}{\bf D}^{-1} {{\bf s}_\text{E}^m} + \left ( {{\bf c}_\text{E}^n} \right )^\text{T}{\bf D}^{-1} {{\bf c}_\text{E}^l} \left ( {{\bf c}_\text{E}^l} \right )^\text{T}{\bf D}^{-1} {{\bf s}_\text{E}^m} \right ] {\bf A}^m \left ( {\bf B}^n \right )^\text{T} \\
 &{\kern 2pt} + \sum_{l=1}^{N} \left [ \left ( {{\bf s}_\text{E}^n} \right )^\text{T}{\bf D}^{-1} {{\bf s}_\text{E}^l} \left ( {{\bf s}_\text{E}^l} \right )^\text{T}{\bf D}^{-1} {{\bf c}_\text{E}^m} + \left ( {{\bf s}_\text{E}^n} \right )^\text{T}{\bf D}^{-1} {{\bf c}_\text{E}^l} \left ( {{\bf c}_\text{E}^l} \right )^\text{T}{\bf D}^{-1} {{\bf c}_\text{E}^m} \right ] {\bf B}^m \left ( {\bf A}^n \right )^\text{T} \\
 &{\kern 2pt} + \sum_{l=1}^{N} \left [ \left ( {{\bf c}_\text{E}^n} \right )^\text{T}{\bf D}^{-1} {{\bf s}_\text{E}^l} \left ( {{\bf s}_\text{E}^l} \right )^\text{T}{\bf D}^{-1} {{\bf c}_\text{E}^m} + \left ( {{\bf c}_\text{E}^n} \right )^\text{T}{\bf D}^{-1} {{\bf c}_\text{E}^l} \left ( {{\bf c}_\text{E}^l} \right )^\text{T}{\bf D}^{-1} {{\bf c}_\text{E}^m} \right ] {\bf B}^m \left ( {\bf B}^n \right )^\text{T} \\
 = &{\kern 2pt} \left ( {{\bf s}_\text{E}^n} \right )^\text{T}{\bf D}^{-1} \underbrace{\left \{ \sum_{l=1}^{N}\left [ {{\bf s}_\text{E}^l} \left ( {{\bf s}_\text{E}^l} \right )^\text{T} + {{\bf c}_\text{E}^l} \left ( {{\bf c}_\text{E}^l} \right )^\text{T} \right ] \right \}}_{\bf D}  {\bf D}^{-1} {{\bf s}_\text{E}^m} {\bf A}^m \left ( {\bf A}^n \right )^\text{T} + \left ( {{\bf c}_\text{E}^n} \right )^\text{T}{\bf D}^{-1} \underbrace{\left \{ \sum_{l=1}^{N}\left [ {{\bf s}_\text{E}^l} \left ( {{\bf s}_\text{E}^l} \right )^\text{T} + {{\bf c}_\text{E}^l} \left ( {{\bf c}_\text{E}^l} \right )^\text{T} \right ] \right \}}_{\bf D}  {\bf D}^{-1} {{\bf s}_\text{E}^m} {\bf A}^m \left ( {\bf B}^n \right )^\text{T} \\
 & + \left ( {{\bf s}_\text{E}^n} \right )^\text{T}{\bf D}^{-1} \underbrace{\left \{ \sum_{l=1}^{N}\left [ {{\bf s}_\text{E}^l} \left ( {{\bf s}_\text{E}^l} \right )^\text{T} + {{\bf c}_\text{E}^l} \left ( {{\bf c}_\text{E}^l} \right )^\text{T} \right ] \right \}}_{\bf D} {\bf D}^{-1} {{\bf c}_\text{E}^m} {\bf B}^m \left ( {\bf A}^n \right )^\text{T} + \left ( {{\bf c}_\text{E}^n} \right )^\text{T}{\bf D}^{-1} \underbrace{\left \{ \sum_{l=1}^{N}\left [ {{\bf s}_\text{E}^l} \left ( {{\bf s}_\text{E}^l} \right )^\text{T} + {{\bf c}_\text{E}^l} \left ( {{\bf c}_\text{E}^l} \right )^\text{T} \right ] \right \}}_{\bf D}  {\bf D}^{-1} {{\bf c}_\text{E}^m} {\bf B}^m \left ( {\bf B}^n \right )^\text{T} \\
 = &{\kern 2pt} \left ( {{\bf s}_\text{E}^n} \right )^\text{T} {\bf D}^{-1} {{\bf s}_\text{E}^m} {\bf A}^m \left ( {\bf A}^n \right )^\text{T} + \left ( {{\bf c}_\text{E}^n} \right )^\text{T}{\bf D}^{-1} {{\bf s}_\text{E}^m} {\bf A}^m \left ( {\bf B}^n \right )^\text{T} + \left ( {{\bf s}_\text{E}^n} \right )^\text{T}{\bf D}^{-1} {{\bf c}_\text{E}^m} {\bf B}^m \left ( {\bf A}^n \right )^\text{T} + \left ( {{\bf c}_\text{E}^n} \right )^\text{T} {\bf D}^{-1} {{\bf c}_\text{E}^m} {\bf B}^m \left ( {\bf B}^n \right )^\text{T} \\
 = &{\kern 2pt} p_{m,n} {\bf A}^m \left ( {\bf A}^n \right )^\text{T} + f_{m,n} {\bf A}^m \left ( {\bf B}^n \right )^\text{T} + g_{m,n} {\bf B}^m \left ( {\bf A}^n \right )^\text{T} + q_{m,n} {\bf B}^m \left ( {\bf B}^n \right )^\text{T} \\
 = &{\kern 2pt} {\bf U}_{m,n} \\
 =&{\kern 2pt}  {\bf U}\left( {m,n} \right)
\label{eq_59}
\end{aligned}
\end{equation}

\hrulefill
\vspace*{4pt}
\end{figure*}

\section{Conclusion}
In this paper, a CI-based block-level precoding algorithm is proposed for the downlink of a MU-MISO communication system. As opposed to traditional CI-based precoding schemes that employ a symbol-level design, the proposed CI-BLP applies a constant precoding matrix to a block of symbol slots within the channel coherence interval, thus greatly reducing the number of optimization problems that need to be solved for CI-based precoding. For both PSK and QAM modulation, the optimal precoding matrix for CI-BLP is derived by constructing the Lagrangian and formulating the KKT conditions. Further manipulations of the dual problem demonstrate that the CI-BLP problem is equivalent to a QP optimization. Numerical results show that owing to the relaxed block-level power constraint, the proposed CI-BLP approach offers an improved performance over the traditional CI-SLP scheme when the length of the considered block is short and the SNR is sufficiently high, and exhibits only a slight performance loss as the block length increases.

\appendices
\section{Construction of ${\bf M}^n$ in \eqref{eq_4}}
Based on Section IV-A of \cite{r25}, ${\bf M}^n \in {\mathbb R}^{2K \times 2N_\text{T}}$ in \eqref{eq_4} can be constructed based on the channel vector ${\bf h}_k$ and the data symbol $s_k^n$ for each user, given by
\begin{equation}
\setcounter{equation}{56}
{\bf M}^n=\begin{bmatrix}
 {\bf j}_1 & {\bf j}_2 & \cdots & {\bf j}_K & {\bf l}_1 & {\bf l}_2 & \cdots & {\bf l}_K
\end{bmatrix}^\text{T},
\label{eq_56}
\end{equation}
where ${\bf j}_k \in {\mathbb R}^{2N_\text{T}}$ and ${\bf l}_k \in {\mathbb R}^{2N_\text{T}}$ are given by
\begin{equation}
{\bf j}_k=
\begin{bmatrix}
 \frac{\Im\left ( s_{k, {\cal B}}^n \right ) \Re\left ( {\bf h}_k \right )-\Re \left ( s_{k, {\cal B}}^n \right ) \Im\left ( {\bf h}_k \right ) }{\Re\left ( s_{k, {\cal A}}^n \right )\Im\left ( s_{k, {\cal B}}^n \right )-\Im\left ( s_{k, {\cal A}}^n \right )\Re\left ( s_{k, {\cal B}}^n \right )}
\\
\\
 -\frac{\Im\left ( s_{k, {\cal B}}^n \right ) \Im\left ( {\bf h}_k \right )+\Re \left ( s_{k, {\cal B}}^n \right ) \Re\left ( {\bf h}_k \right ) }{\Re\left ( s_{k, {\cal A}}^n \right )\Im\left ( s_{k, {\cal B}}^n \right )-\Im\left ( s_{k, {\cal A}}^n \right )\Re\left ( s_{k, {\cal B}}^n \right )}
\end{bmatrix},
\label{eq_57}
\end{equation}
and
\begin{equation}
{\bf l}_k=\begin{bmatrix}
 \frac{\Re\left ( s_{k, {\cal A}}^n \right ) \Im\left ( {\bf h}_k \right )-\Im \left ( s_{k, {\cal A}}^n \right ) \Re\left ( {\bf h}_k \right ) }{\Re\left ( s_{k, {\cal A}}^n \right )\Im\left ( s_{k, {\cal B}}^n \right )-\Im\left ( s_{k, {\cal A}}^n \right )\Re\left ( s_{k, {\cal B}}^n \right )}
\\
\\
 \frac{\Re\left ( s_{k, {\cal A}}^n \right ) \Re\left ( {\bf h}_k \right )+\Im \left ( s_{k, {\cal A}}^n \right ) \Im\left ( {\bf h}_k \right ) }{\Re\left ( s_{k, {\cal A}}^n \right )\Im\left ( s_{k, {\cal B}}^n \right )-\Im\left ( s_{k, {\cal A}}^n \right )\Re\left ( s_{k, {\cal B}}^n \right )}
\end{bmatrix}.
\label{eq_58}
\end{equation}

\section{Proof for Proposition 1}
We begin by expressing the generic $(m,n)$-th block in $\left( {\bf F} + {\bf G} \right)$, which is shown in \eqref{eq_59} on the top of this page based on (30) and \eqref{eq_34}, where we obtain that
\begin{equation}
\setcounter{equation}{60}
{\bf F}_{m,n} + {\bf G}_{m,n} = {\bf U}_{m,n}, {\kern 3pt} \forall m,n \in {\cal N},
\label{eq_60}
\end{equation}
which completes the proof. $\blacksquare$

\ifCLASSOPTIONcaptionsoff
  \newpage
\fi

\bibliographystyle{IEEEtran}
\bibliography{refs.bib}

\end{document}